\documentclass[reprint,showpacs,groupedaddress,amsfonts,amsmath,amssymb,aps,prd]{revtex4-1}
\usepackage{graphicx}
\usepackage{dcolumn}% Align table columns on decimal point
\usepackage{bm}% bold math
\begin{document}
\preprint{APS/123-QED}
\title{The spherically symmetric solution of $f(R,\mathcal{G})$ gravity at low energy}

\newcommand*{\PKU}{School of Physics and State Key Laboratory of Nuclear Physics and
Technology, Peking University, Beijing 100871,
China}\affiliation{\PKU}
\newcommand*{\INFN}{INFN, Sez. di Pavia, via Bassi 6, 27100 Pavia, Italy}\affiliation{\INFN}
\newcommand*{\CICQM}{Collaborative Innovation
Center of Quantum Matter, Beijing, China}\affiliation{\CICQM}
\newcommand*{\CHEP}{Center for High Energy
Physics, Peking University, Beijing 100871,
China}\affiliation{\CHEP}

\author{Bofeng Wu}\email{bofengw@pku.edu.cn}\affiliation{\PKU}
\author{Bo-Qiang Ma}\email{mabq@pku.edu.cn}\affiliation{\PKU}\affiliation{\CICQM}\affiliation{\CHEP}

%\date{\today}
\begin{abstract}
The weak-field and slow-motion limit of $f(R,\mathcal{G})$ gravity is developed up to $(v/c)^{4}$ order in spherically symmetric background. Considering the Taylor expansion of a general function $f$ around vanishing values of $R$ and $\mathcal{G}$, we present general vacuum solutions up to $(v/c)^{4}$ order for the gravitational field generated by a ball-like source. The spatial behaviors at $(v/c)^{2}$ order are the same for $f(R,\mathcal{G})$ gravity and $f(R)$ gravity, and their corresponding real valued static behaviors are presented and compared with the one in general relativity. The static Yukawa-like behavior is proved to be compatible with the previous result of the most general fourth-order theory. At $(v/c)^{4}$ order, the static corrections to the Yukawa-like behavior for $f(R,\mathcal{G})$ gravity, $f(R)$ gravity, and the Starobinsky gravity are presented and compared with the one in general relativity.
\end{abstract}
\maketitle

\section{Introduction}
As is well known, General Relativity (GR) is a successful and elegant theory of gravity, but it has many challenges to interpret a growing number of data observed at infrared scales. Among them, the most famous one is that GR has to introduce additional concepts like dark energy/matter to interpret the observed data about the universe, and this is regarded to be the signal of a breakthrough of GR at astrophysical and cosmology scales~\cite{1,2}.

Another approach to deal with the problems mentioned above is to introduce the Extended Theories of Gravity (ETG)~\cite{3,4,5,6}. These theories are based on generalizations and extensions of GR.
$f(R)$ gravity~\cite{7,8,9,10} is a famous theory which modifies the Einstein-Hilbert action by adopting a general function of the Ricci scalar $R$ in the gravitational Lagrangian density so as to explain the inflationary behavior with respect to  the early universe, as shown by Starobinsky~\cite{7}. $f(R)$ gravity is a typical example of higher order gravity (HOG).

The Gauss-Bonnet (GB) curvature invariant $\mathcal{G}$ is another interesting curvature quantity. This term can avoid ghost contributions and contribute to the regularization of the gravitational action~\cite{11,12,13}.  Recently, a new generalized modified the GB gravity, whose Lagrangian density is a general function of $R$ and $\mathcal{G}$ as $f(R,\mathcal{G})$, has attracted considerable attention~\cite{14,15}.

In order to test the viability of ETG and parameterize their deviations with respect to GR, we should take into account their weak-field and slow-motion (WFSM) limit~\cite{16}. ETG usually yield corrections to the Newton potential and the Eddington-parameters which could be a test for these theories~\cite{16,17}. In the last few years, several authors have dealt with the WFSM limit of HOG~\cite{18,19,20}. These papers shed new light on the WFSM limit of such theories.

In Refs.~\cite{21,22}, Capozziello, Stabile, and Troisi built a new formulism to deal with the Newtonian limit of $f(R)$ gravity in the spherically symmetric background with the metric approach in the Jordan frame. In this formulism, the differential equations at $(v/c)^{4}$ order for $f(R)$ gravity are so involuted that corresponding general vacuum solutions are not obtained easily. In fact, in order to obtain further results, Stabile adopted the system of isotropic coordinates~\cite{16} to study the post-Newtonian limit of $f(R)$ gravity~\cite{23} and the Newtonian limit of fourth order gravity~\cite{24,25}. In addition, in Ref.~\cite{26} the differential equations of $f(R,\mathcal{G})$ gravity are calculated up to $(v/c)^{6}$ order by using the system of isotropic coordinates, and
the solution in the Newtonian limit is obtained.

In this paper, we develop the WFSM limit of $f(R,\mathcal{G})$ gravity in spherically symmetric background up to $(v/c)^{4}$ order by generalizing the formalism mentioned above~\cite{21,22}. Considering the Taylor expansion of a general function $f$ around vanishing values of $R$ and $\mathcal{G}$, we present general vacuum solutions up to $(v/c)^{4}$ order for $f(R,\mathcal{G})$ gravity in a pure perturbative framework. These solutions are time-dependent, and the time-dependent evolution depends on the order of perturbations. At $(v/c)^{2}$ order, compared with the conclusions in Refs.~\cite{21,22}, we find that both $f(R,\mathcal{G})$ gravity and $f(R)$ gravity have the same spatial behaviors, namely the Yukawa-like behavior and the oscillating-like behavior, and the latter is complex valued and does not meet asymptotic condition in general. For the gravitational field generated by a ball-like source, we show that these two behaviors related to the $g_{tt}$ components provide two kinds of corrected gravitational potentials about the Newtonian one. Furthermore for such gravitational field, we present its two real valued static behaviors and compare them with the one in GR.

In Ref.~\cite{25}, the Newtonian limit of the most general fourth-order theory of gravity, namely $F(X,Y,Z)$ gravity, has been studied, with $X=R$, $Y=R_{\mu\nu}R^{\mu\nu}$, and $Z=R_{\mu\nu\sigma\rho}R^{\mu\nu\sigma\rho}$. According to  its conclusions, for the gravitational field generated by a ball-like source, its static Yukawa-like behavior has two characteristic lengths $m_{1}^{-1}$ and $m_{2}^{-1}$. $m_{1}$ and $m_{2}$ are defined as
\begin{equation}\label{1}
m_{1}=\left(-\frac{F_{X}(0,0,0)}{3F_{XX}(0,0,0)+2F_{Y}(0,0,0)+2F_{Z}(0,0,0)}\right)^{\frac{1}{2}}
\end{equation}
and
\begin{equation}\label{2}
m_{2}=\left(\frac{F_{X}(0,0,0)}{F_{Y}(0,0,0)+4F_{Z}(0,0,0)}\right)^{\frac{1}{2}},
\end{equation}
where $F_{X}=\frac{\partial F}{\partial X}$, $F_{Y}=\frac{\partial F}{\partial Y}$, and $F_{Z}=\frac{\partial F}{\partial Z}$. But in our paper, for $f(R,\mathcal{G})$ gravity, we show that its static Yukawa-like behavior has only one characteristic length, and this is the consequence of
\begin{equation}\label{3}
\mathcal{G}=X^{2}-4Y+Z.
\end{equation}
Thus, our result about the Yukawa-like behavior is compatible with the corresponding one in $F(X,Y,Z)$ gravity. In addition for such gravitational field, we find that the gravitational potential has the divergency at the position of source, and this conclusion is different from corresponding one of $F(X,Y,Z)$ gravity in Ref.~\cite{25}.

At $(v/c)^{3}$ order, although the differential equation is related to the time $t$, we find that the above two behaviors at $(v/c)^{2}$ order meet this equation, so that we can not fix the time-dependent evolution of these two behaviors at this order. In fact, the differential equations at $(v/c)^{4}$ order still can not solve this problem, and this implies that in order to probe the time-dependent evolution about the behaviors at $(v/c)^{2}$ order, we need to develop the WFSM limit of $f(R,\mathcal{G})$ gravity up to more orders.

The correction to the Yukawa-like behavior at $(v/c)^{2}$ order is calculated to $(v/c)^{4}$ order in our paper, and by this correction and previous results, we draw a conclusion: If the gravitational field is generated by a ball-like source, for the submodel of $f(R)$ gravity whose term $R^2$ disappears in the Taylor expansion of $f$ around a vanishing value of $R$, its general vacuum solutions up to $(v/c)^{4}$ order in spherically symmetric background are the same with those in GR. At last, we present the static corrections of such gravitational field to the Yukawa-like behavior for $f(R,\mathcal{G})$ gravity, $f(R)$ gravity, and the Starobinsky gravity~\cite{27}, and compare them with the one in GR.

This paper is organized as follows. In Sec.$\ $\uppercase\expandafter{\romannumeral2}, we review the equation of $f(R,\mathcal{G})$ gravity. In Sec.$\ $\uppercase\expandafter{\romannumeral3},
We report the complete scheme of the WFSM limit for $f(R,\mathcal{G})$ gravity up to $(v/c)^{4}$ order in spherically symmetric background, and the then obtain the corresponding general vacuum solutions. General comments with respect to the mathematical properties of differential equations, their relative solutions, and the asymptotic behavior of the metric tensor are reported. In Sec.$\ $\uppercase\expandafter{\romannumeral4}, we summarize the obtained results.
\section{Field equation of $f(R,\mathcal{G})$ gravity}
In the following of this work, we use the unit $c=1$.
The starting action of $f(R,\mathcal{G})$ gravity is
\begin{equation}\label{4}
S=\frac{1}{2\kappa}\int dx^4\sqrt{-g}f(R,\mathcal{G})+S_{M}(g^{\mu\nu},\psi),
\end{equation}
where $S_{M}(g^{\mu\nu},\psi)$ is the matter action, $g$ is the determinant of metric, and $\kappa=8\pi G$. The GB invariant is defined by (\ref{3}). The gravitational field equation~\cite{28} of $f(R,\mathcal{G})$  gravity and corresponding trace equation~\cite{26} are respectively
\begin{equation}\label{5}
H_{\mu\nu}=\kappa T_{\mu\nu},\qquad H=\kappa T,
\end{equation}
where
\begin{widetext}
\begin{align*}
H_{\mu\nu}=&-\frac{g_{\mu\nu}}{2}f+R_{\mu\nu}f_{R}+g_{\mu\nu}\square f_{R}-\triangledown_{\mu}\triangledown_{\nu} f_{R}+\frac{g_{\mu\nu}}{2}f_{\mathcal{G}}\mathcal{G}+2R g_{\mu\nu}\square f_{\mathcal{G}}-2R\triangledown_{\mu}\triangledown_{\nu}f_{\mathcal{G}}
\label{6}\\&+4R_{\nu}^{\phantom{\nu}\lambda}\triangledown_{\lambda}\triangledown_{\mu}f_{\mathcal{G}}+
4R_{\mu}^{\phantom{\mu}\lambda}\triangledown_{\lambda}\triangledown_{\nu}f_{\mathcal{G}}-4g_{\mu\nu}R^{\alpha\beta}
\triangledown_{\alpha}\triangledown_{\beta}f_{\mathcal{G}}-4R_{\mu\nu}\square f_{\mathcal{G}}+4R_{\mu\rho\nu\sigma}\triangledown^{\rho}\triangledown^{\sigma}f_{\mathcal{G}},\tag{6}\\
\label{7}H=&-2f+R f_{R}+3\square f_{R}+2f_{\mathcal{G}}\mathcal{G}+2R\square f_{\mathcal{G}}-4R^{\alpha\beta}\triangledown_{\alpha}\triangledown_{\beta}f_{\mathcal{G}},\tag{7}
\end{align*}
\end{widetext}
\begin{displaymath}
f_{R}=\frac{\partial f}{\partial R},\qquad f_{\mathcal{G}}=\frac{\partial f}{\partial \mathcal{G}},
\end{displaymath}
and $T_{\mu\nu}$ is the energy-momentum tensor describing the ordinary matter.

\section{The WFSM limit up to $O(4)$ order of $f(R,\mathcal{G})$ gravity in spherically symmetric background}
Here we do not discuss the theory on how to formulate a mathematically well founded WFSM limit of some theory about gravity in spherically symmetric background, but we recommend Refs.~\cite{21,22} to the interested readers. We provide the explicit form of all the quantities involved in the WFSM limit in spherically symmetric background for $f(R,\mathcal{G})$ gravity.

In the solar system, all the quantities involved in the WFSM formulism can be expanded in powers of $\overline{v}^2$, where $\overline{v}$ is planetary average velocities, and is small with respect to the light speed.
Moreover for typical potential energy $U$, the matter pressure $p$, and the matter density $\rho$, they satisfy the relationship
\begin{equation}\label{8}
U\sim \frac{p}{\rho}\sim\overline{v}^2\sim O(2),\tag{8}
\end{equation}
and then one has
\begin{equation}\label{9}
\frac{|\partial/\partial x^0|}{|\nabla|}\sim O(1),\tag{9}
\end{equation}
where $x^0=t$.

We assume a spherically symmetric spacetime whose metric tensor is
\begin{equation}\label{10}
g_{\mu\nu}=\left(\begin{array}{cccc}g_{tt}(t,r)&0&0&0\\0&g_{rr}(t,r)&0&0\\0&0&-r^2&0\\0&0&0&-r^2\sin^2\theta\end{array}\right).\tag{10}
\end{equation}
\subsection{The expansion of all involved quantities in powers of $\overline{v}^2$}
In order to build the WSLM limit of $f(R,\mathcal{G})$ gravity and give its general solution, for the given $g_{\mu\nu}$ in (\ref{10}), we assume
\begin{equation}\label{11}
\left\{\begin{array}{l}
g_{tt}(t,r)\simeq 1+g_{tt}^{(2)}(t,r)+g_{tt}^{(4)}(t,r),\\
g_{rr}(t,r)\simeq -1+g_{rr}^{(2)}(t,r)+g_{rr}^{(4)}(t,r),\\
g_{\theta\theta}(t,r)= -r^2,\\
g_{\varphi\varphi}(t,r)= -r^2\sin^2\theta
\end{array}\right.\tag{11}
\end{equation}
and \begin{equation}\label{12}
\left\{\begin{array}{l}
g^{tt}(t,r)\simeq 1+g^{(2)tt}(t,r)+g^{(4)tt}(t,r),\\
g^{rr}(t,r)\simeq -1+g^{(2)rr}(t,r)+g^{(4)rr}(t,r),\\
g^{\theta\theta}(t,r)=-\frac{1}{r^2},\\
g^{\varphi\varphi}(t,r)=-\frac{1}{r^2\sin^2\theta},
\end{array}\right.\tag{12}
\end{equation}
and then by using the equalities $g^{tt}g_{tt}=g^{rr}g_{rr}=1$, there are
\begin{equation}\label{13}
\left\{\begin{array}{ll}
g^{(2)tt}=-g_{tt}^{(2)},&g^{(2)rr}=-g_{rr}^{(2)},\\
g^{(4)tt}=-g_{tt}^{(4)}+g_{tt}^{(2)2},&g^{(4)rr}=-g_{rr}^{(4)}-g_{rr}^{(2)2}.
\end{array}
\right.\tag{13}
\end{equation}
By assumptions (\ref{11}) and (\ref{12}), the orders involved in this ansatz for metric tensor are beyond the post-Newtonian order, and that is what we are mainly interested in this work.

By the metric tensor $g_{\mu\nu}$, the associated connection can be derived as
\begin{displaymath}
\Gamma^{\alpha}_{\mu\nu}=\frac{1}{2}g^{\alpha\beta}(\partial_{\mu}g_{\nu\beta}+\partial_{\nu}g_{\mu\beta}-\partial_{\beta}g_{\mu\nu}),
\end{displaymath}
and then by (\ref{10})--(\ref{13}), there are
\begin{equation}\label{14}
\left\{
\begin{array}{l}
\Gamma^{t}_{tt}=\Gamma^{(3)t}_{\phantom{(3)}tt}+O(5),\\
\Gamma^{r}_{tt}=\Gamma^{(2)r}_{\phantom{(3)}tt}+\Gamma^{(4)r}_{\phantom{(3)}tt}+O(6),\\
\Gamma^{t}_{tr}=\Gamma^{(2)t}_{\phantom{(3)}tr}+\Gamma^{(4)t}_{\phantom{(3)}tr}+O(6),\\
\Gamma^{r}_{tr}=\Gamma^{(3)r}_{\phantom{(3)}tr}+O(5),\\
\Gamma^{t}_{rr}=\Gamma^{(3)t}_{\phantom{(3)}rr}+O(5),\\
\Gamma^{r}_{rr}=\Gamma^{(2)r}_{\phantom{(3)}rr}+\Gamma^{(4)r}_{\phantom{(3)}rr}+O(6),\\
\Gamma^{r}_{\theta\theta}=\Gamma^{(0)r}_{\phantom{(3)}\theta\theta}+\Gamma^{(2)r}_{\phantom{(3)}\theta\theta}+
\Gamma^{(4)r}_{\phantom{(3)}\theta\theta}+O(6),\\
\Gamma^{r}_{\varphi\varphi}=\Gamma^{r}_{\theta\theta}\sin^2\theta,\\
\Gamma^{\theta}_{r\theta}=\Gamma^{(0)\theta}_{\phantom{(3)}r\theta}=\Gamma^{\varphi}_{r\varphi}=\Gamma^{(0)\varphi}_{\phantom{(3)}r\varphi}=\frac{1}{r},\\
\Gamma^{\theta}_{\varphi\varphi}=\Gamma^{(0)\theta}_{\phantom{(3)}\varphi\varphi}=-\sin\theta\cos\theta,\\
\Gamma^{\varphi}_{\theta\varphi}=\cot\theta,
\end{array}\right.\tag{14}
\end{equation}
where
\begin{equation}\label{15}
\left\{
\begin{array}{l}
\Gamma^{(3)t}_{\phantom{(3)}tt}=\frac{1}{2}g_{tt,t}^{(2)},\\
\Gamma^{(2)r}_{\phantom{(3)}tt}=\frac{1}{2}g_{tt,r}^{(2)},\\
\Gamma^{(4)r}_{\phantom{(3)}tt}=\frac{1}{2}g_{tt,r}^{(4)}+\frac{1}{2}g_{rr}^{(2)}g_{tt,r}^{(2)},\\
\Gamma^{(2)t}_{\phantom{(3)}tr}=\frac{1}{2}g_{tt,r}^{(2)},\\
\Gamma^{(4)t}_{\phantom{(3)}tr}=\frac{1}{2}g_{tt,r}^{(4)}-\frac{1}{2}g_{tt}^{(2)}g_{tt,r}^{(2)},\\
\Gamma^{(3)r}_{\phantom{(3)}tr}=-\frac{1}{2}g_{rr,t}^{(2)},\\
\Gamma^{(3)t}_{\phantom{(3)}rr}=-\frac{1}{2}g_{rr,t}^{(2)},\\
\Gamma^{(2)r}_{\phantom{(3)}rr}=-\frac{1}{2}g_{rr,r}^{(2)},\\
\Gamma^{(4)r}_{\phantom{(3)}rr}=-\frac{1}{2}g_{rr,r}^{(4)}-\frac{1}{2}g_{rr}^{(2)}g_{rr,r}^{(2)},\\
\Gamma^{(0)r}_{\phantom{(3)}\theta\theta}=-r,\\
\Gamma^{(2)r}_{\phantom{(3)}\theta\theta}=-r g_{rr}^{(2)},\\
\Gamma^{(4)r}_{\phantom{(3)}\theta\theta}=-r g_{rr}^{(4)}-r g_{rr}^{(2)2}.
\end{array}\right.\tag{15}
\end{equation}

After obtaining the connection, by formula
\begin{displaymath}
R^{\mu}_{\phantom{\mu}\nu\rho\sigma}=\Gamma^{\mu}_{\sigma\nu,\rho}-\Gamma^{\mu}_{\rho\nu,\sigma}
+\Gamma^{\mu}_{\alpha\rho}\Gamma^{\alpha}_{\sigma\nu}-\Gamma^{\mu}_{\alpha\sigma}\Gamma^{\alpha}_{\rho\nu},
\end{displaymath}
the Riemann tensor can be calculated to be
\begin{widetext}
\begin{equation}\label{16}
\left\{\begin{array}{ll}
R^{\mu}_{\phantom{\mu}\nu\rho\sigma}=R^{(2)\mu}_{\phantom{(2)\mu}\nu\rho\sigma}+R^{(4)\mu}_{\phantom{(4)\mu}\nu\rho\sigma}+O(6),&\text{the number of letter ``t" in set $\{\mu,\nu,\rho,\sigma\}$ is even,}\\
R^{\mu}_{\phantom{\mu}\nu\rho\sigma}=R^{(3)\mu}_{\phantom{(2)\mu}\nu\rho\sigma}+O(5),&\text{the number of letter ``t" in set $\{\mu,\nu,\rho,\sigma\}$ is odd,}
\end{array}\right.\tag{16}
\end{equation}
\end{widetext}
where
\begin{equation}\label{17}
\left\{\begin{array}{l}
R^{(2)r}_{\phantom{(3r)}trt}=\frac{1}{2}g^{(2)}_{tt,rr},\\
R^{(4)r}_{\phantom{(3r)}trt}=\frac{1}{2}g^{(4)}_{tt,rr}+\frac{1}{2}g^{(2)}_{rr}g^{(2)}_{tt,rr}\\
\qquad\qquad\ +\frac{1}{4}g^{(2)}_{rr,r}g^{(2)}_{tt,r}-\frac{1}{4}g^{(2)2}_{tt,r}\\
\qquad\qquad\ +\frac{1}{2}g^{(2)}_{rr,tt},\\
R^{(2)t}_{\phantom{(3r)}rrt}=\frac{1}{2}g^{(2)}_{tt,rr},\\
R^{(4)t}_{\phantom{(3r)}rrt}=\frac{1}{2}g^{(4)}_{tt,rr}-\frac{1}{2}g^{(2)}_{tt}g^{(2)}_{tt,rr}\\
\qquad\qquad\ +\frac{1}{4}g^{(2)}_{rr,r}g^{(2)}_{tt,r}-\frac{1}{4}g^{(2)2}_{tt,r}\\
\qquad\qquad\ +\frac{1}{2}g^{(2)}_{rr,tt},\\
R^{(2)\theta}_{\phantom{(3r)}t\theta t}=\frac{1}{2r}g^{(2)}_{tt,r},\\
R^{(4)\theta}_{\phantom{(3r)}t\theta t}=\frac{1}{2r}g^{(4)}_{tt,r}+\frac{1}{2r}g^{(2)}_{rr}g^{(2)}_{tt,r},\\
R^{(3)\theta}_{\phantom{(3r)}t\theta r}=-\frac{1}{2r}g^{(2)}_{rr,t},\\
R^{(3)\theta}_{\phantom{(3r)}r\theta t}=-\frac{1}{2r}g^{(2)}_{rr,t},\\
R^{(2)\theta}_{\phantom{(3r)}r\theta r}=-\frac{1}{2r}g^{(2)}_{rr,r},\\
R^{(4)\theta}_{\phantom{(3r)}r\theta r}=-\frac{1}{2r}g^{(4)}_{rr,r}-\frac{1}{2r}g^{(2)}_{rr}g^{(2)}_{rr,r},\\
R^{(2)t}_{\phantom{(3r)}\theta\theta t}=\frac{r}{2}g^{(2)}_{tt,r},\\
R^{(4)t}_{\phantom{(3r)}\theta\theta t}=\frac{r}{2}g^{(4)}_{tt,r}-\frac{r}{2}g^{(2)}_{tt}g^{(2)}_{tt,r}\\
\qquad\qquad\ \ +\frac{r}{2}g^{(2)}_{rr}g^{(2)}_{tt,r},\\
R^{(3)r}_{\phantom{(3r)}\theta\theta t}=\frac{r}{2}g^{(2)}_{rr,t},\\
R^{(3)t}_{\phantom{(3r)}\theta\theta r}=-\frac{r}{2}g^{(2)}_{rr,t},\\
R^{(2)r}_{\phantom{(3r)}\theta\theta r}=\frac{r}{2}g^{(2)}_{rr,r},\\
R^{(4)r}_{\phantom{(3r)}\theta\theta r}=\frac{r}{2}g^{(4)}_{rr,r}+r g^{(2)}_{rr}g^{(2)}_{rr,r},\\
R^{(2)\varphi}_{\phantom{(3r)}\theta\varphi\theta}=-g^{(2)}_{rr},\\
R^{(4)\varphi}_{\phantom{(3r)}\theta\varphi \theta}=-g^{(4)}_{rr}-g^{(2)2}_{rr}
\end{array}\right.\tag{17}
\end{equation}
and
\begin{equation}\label{18}
\left\{\begin{array}{l}
R^{\varphi}_{\phantom{\varphi}t\varphi t}=R^{\theta}_{\phantom{\theta}t\theta t},\\
R^{\varphi}_{\phantom{\varphi}t\varphi r}=R^{\theta}_{\phantom{\theta}t\theta r},\\
R^{\varphi}_{\phantom{\varphi}r\varphi t}=R^{\theta}_{\phantom{\theta}r\theta t},\\
R^{\varphi}_{\phantom{\varphi}r\varphi r}=R^{\theta}_{\phantom{\theta}r\theta r},\\
R^{t}_{\phantom{r}\varphi\varphi t}=R^{t}_{\phantom{t}\theta\theta t}\sin^2\theta,\\
R^{r}_{\phantom{r}\varphi\varphi t}=R^{r}_{\phantom{r}\theta\theta t}\sin^2\theta,\\
R^{t}_{\phantom{r}\varphi\varphi r}=R^{t}_{\phantom{r}\theta\theta r}\sin^2\theta,\\
R^{r}_{\phantom{r}\varphi\varphi r}=R^{r}_{\phantom{r}\theta\theta r}\sin^2\theta,\\
R^{\theta}_{\phantom{\theta}\varphi\varphi \theta}=-R^{\varphi}_{\phantom{\varphi}\theta\varphi\theta}\sin^2\theta.
\end{array}\right.\tag{18}
\end{equation}
Then, using formulas $R_{\mu\nu}=R^{\alpha}_{\phantom{\alpha}\mu\alpha\nu}$, $R=g^{\mu\nu}R_{\mu\nu}$, and equations in (\ref{13}), we obtain the Ricci tensor and the Ricci scalar
\begin{equation}\label{19}
\left\{\begin{array}{l}
R_{tt}=R^{(2)}_{tt}+R^{(4)}_{tt}+O(6),\\
R_{rr}=R^{(2)}_{rr}+R^{(4)}_{rr}+O(6),\\
R_{\theta\theta}=R^{(2)}_{\theta\theta}+R^{(4)}_{\theta\theta}+O(6),\\
R_{\varphi\varphi}=R^{(2)}_{\varphi\varphi}+R^{(4)}_{\varphi\varphi}+O(6),\\
R_{tr}=R^{(3)}_{tr}+O(5),\\
R=R^{(2)}+R^{(4)}+O(6)
\end{array}\right.\tag{19}
\end{equation}
with
\begin{equation}\label{20}
\left\{\begin{array}{l}
R^{(2)}_{tt}=\frac{1}{2}g^{(2)}_{tt,rr}+\frac{1}{r}g^{(2)}_{tt,r},\\
R^{(4)}_{tt}=\frac{1}{2}g^{(4)}_{tt,rr}+\frac{1}{2}g^{(2)}_{rr}g^{(2)}_{tt,rr}
+\frac{1}{4}g^{(2)}_{rr,r}g^{(2)}_{tt,r}\\
\qquad\quad-\frac{1}{4}g^{(2)2}_{tt,r}+\frac{1}{2}g^{(2)}_{rr,tt}
+\frac{1}{r}g^{(4)}_{tt,r}+\frac{1}{r}g^{(2)}_{rr}g^{(2)}_{tt,r},\\
R^{(2)}_{rr}=-\frac{1}{2}g^{(2)}_{tt,rr}-\frac{1}{r}g^{(2)}_{rr,r},\\
R^{(4)}_{rr}=-\frac{1}{2}g^{(4)}_{tt,rr}+\frac{1}{2}g^{(2)}_{tt}g^{(2)}_{tt,rr}
-\frac{1}{4}g^{(2)}_{rr,r}g^{(2)}_{tt,r}\\
\qquad\quad+\frac{1}{4}g^{(2)2}_{tt,r}-\frac{1}{2}g^{(2)}_{rr,tt}
-\frac{1}{r}g^{(4)}_{rr,r}-\frac{1}{r}g^{(2)}_{rr}g^{(2)}_{rr,r},\\
R^{(2)}_{\theta\theta}=-g^{(2)}_{rr}-\frac{r}{2}g^{(2)}_{tt,r}-\frac{r}{2}g^{(2)}_{rr,r},\\
R^{(4)}_{\theta\theta}=-g^{(4)}_{rr}-g^{(2)2}_{rr}-\frac{r}{2}g^{(4)}_{tt,r}+\frac{r}{2}g^{(2)}_{tt}g^{(2)}_{tt,r}\\
\qquad\quad-\frac{r}{2}g^{(2)}_{rr}g^{(2)}_{tt,r}-\frac{r}{2}g^{(4)}_{rr,r}-r g^{(2)}_{rr}g^{(2)}_{rr,r},\\
R_{\varphi\varphi}=R_{\theta\theta}\sin^2\theta,\\
R^{(3)}_{tr}=-\frac{1}{r}g^{(2)}_{rr,t},\\
R^{(2)}=g^{(2)}_{tt,rr}+\frac{2}{r^2}g^{(2)}_{rr}+\frac{2}{r}g^{(2)}_{tt,r}+\frac{2}{r}g^{(2)}_{rr,r},\\
R^{(4)}=g^{(4)}_{tt,rr}+\frac{2}{r^2}g^{(4)}_{rr}+\frac{2}{r}g^{(4)}_{tt,r}+\frac{2}{r}g^{(4)}_{rr,r}\\
\qquad\quad-g^{(2)}_{tt}g^{(2)}_{tt,rr}+g^{(2)}_{rr}g^{(2)}_{tt,rr}+\frac{2}{r^2}g^{(2)2}_{rr}\\
\qquad\quad-\frac{2}{r}g^{(2)}_{tt}g^{(2)}_{tt,r}+\frac{2}{r}g^{(2)}_{rr}g^{(2)}_{tt,r}+\frac{4}{r}g^{(2)}_{rr}g^{(2)}_{rr,r}\\
\qquad\quad+\frac{1}{2}g^{(2)}_{tt,r}g^{(2)}_{rr,r}-\frac{1}{2}g^{(2)2}_{tt,r}+g^{(2)}_{rr,tt}.
\end{array}\right.\tag{20}
\end{equation}
Finally by the definition of $\mathcal{G}$ (\ref{3}) and (\ref{13})--(\ref{20}), the expansion of $\mathcal{G}$ in powers of $\overline{v}^2$ can be calculated to be
\begin{equation}\label{21}
\mathcal{G}=\mathcal{G}^{(4)}+O(6),\tag{21}
\end{equation}
where
\begin{equation}\label{22}
\mathcal{G}^{(4)}=\frac{4}{r^2}g^{(2)}_{rr}g^{(2)}_{tt,rr}+\frac{4}{r^2}g^{(2)}_{tt,r}g^{(2)}_{rr,r}.\tag{22}
\end{equation}

On the matter side, we consider a ball-like source with mass $M$ and radius $\xi$, and the corresponding energy-momentum tensor is
(we are not interesting in the internal structure)~\cite{23}
\begin{equation}\label{23}
T_{\mu\nu}=\rho(r)u_{\mu}u_{v},\quad T=\rho(r),\tag{23}
\end{equation}
where
\begin{equation}\label{24}
\rho(r)=\frac{3M}{4\pi \xi^{3}},\quad g^{tt}u_{t}^{2}=1.\tag{24}
\end{equation}

In order to expand the field equation and its trace equation, we take into account the Taylor expansion of $f$ around vanishing values $R$ and $\mathcal{G}$, namely
\begin{align*}
&f(R,\mathcal{G})=f_{0}+f_{1}R+f_{2}\mathcal{G}+\frac{1}{2}(f_{11}R^2+2f_{12}R\mathcal{G}+f_{22}\mathcal{G}^2)\\
\label{25}&+\frac{1}{6}(f_{111}R^3+3f_{112}R^2\mathcal{G}+3f_{122}R\mathcal{G}^2+f_{222}\mathcal{G}^3)+\cdots,\tag{25}
\end{align*}
where
\begin{equation}\label{26}
\left\{\begin{array}{lll}
f_{0}=f(0,0),&f_{1}=f_{R}(0,0),&f_{2}=f_{\mathcal{G}}(0,0),\\
f_{11}=f_{RR}(0,0),&f_{12}=f_{R\mathcal{G}}(0,0),&f_{22}=f_{\mathcal{G}\mathcal{G}}(0,0),\\
f_{111}=f_{RRR}(0,0),&f_{112}=f_{RR\mathcal{G}}(0,0),&f_{122}=f_{R\mathcal{G}\mathcal{G}}(0,0),\\
f_{222}=f_{\mathcal{G}\mathcal{G}\mathcal{G}}(0,0).&&\tag{26}
\end{array}\right.
\end{equation}
By (\ref{10})--(\ref{26}), we obtain
\begin{equation}\label{27}
\left\{\begin{array}{l}
f=f^{(0)}+f^{(2)}+f^{(4)}+O(6),\\
f_{R}=f_{R}^{(0)}+f_{R}^{(2)}+f_{R}^{(4)}+O(6),\\
f_{\mathcal{G}}=f_{\mathcal{G}}^{(0)}+f_{\mathcal{G}}^{(2)}+f_{\mathcal{G}}^{(4)}+O(6),\\
f_{RR}=f_{RR}^{(0)}+f_{RR}^{(2)}+O(4),\\
f_{R\mathcal{G}}=f_{R\mathcal{G}}^{(0)}+f_{R\mathcal{G}}^{(2)}+O(4),\\
f_{\mathcal{G}\mathcal{G}}=f_{\mathcal{G}\mathcal{G}}^{(0)}+f_{\mathcal{G}\mathcal{G}}^{(2)}+O(4),\\
f_{RRR}=f_{RRR}^{(0)}+O(2),\\
f_{RR\mathcal{G}}=f_{RR\mathcal{G}}^{(0)}+O(2),\\
f_{R\mathcal{G}\mathcal{G}}=f_{R\mathcal{G}\mathcal{G}}^{(0)}+O(2),\\
f_{\mathcal{G}\mathcal{G}\mathcal{G}}=f_{\mathcal{G}\mathcal{G}\mathcal{G}}^{(0)}+O(2),
\end{array}\right.\tag{27}
\end{equation}
where
\begin{equation}\label{28}
\left\{\begin{array}{l}
f^{(0)}=f_{0},\qquad\qquad f^{(2)}=f_{1}R^{(2)},\\
f^{(4)}=f_{1}R^{(4)}+f_{2}\mathcal{G}^{(4)}+\frac{f_{11}}{2}R^{(2)2},\\
f_{R}^{(0)}=f_{1},\qquad\qquad f_{R}^{(2)}=f_{11}R^{(2)},\\
f_{R}^{(4)}=f_{11}R^{(4)}+f_{12}\mathcal{G}^{(4)}+\frac{f_{111}}{2}R^{(2)2},\\
f_{\mathcal{G}}^{(0)}=f_{2},\qquad\qquad f_{\mathcal{G}}^{(2)}=f_{12}R^{(2)},\\
f_{\mathcal{G}}^{(4)}=f_{12}R^{(4)}+f_{22}\mathcal{G}^{(4)}+\frac{f_{112}}{2}R^{(2)2},\\
f_{RR}^{(0)}=f_{11},\qquad\quad f_{RR}^{(2)}=f_{111}R^{(2)},\\
f_{R\mathcal{G}}^{(0)}=f_{12},\qquad\quad f_{R\mathcal{G}}^{(2)}=f_{112}R^{(2)},\\
f_{\mathcal{G}\mathcal{G}}^{(0)}=f_{22},\qquad\quad f_{\mathcal{G}\mathcal{G}}^{(2)}=f_{122}R^{(2)},\\
f_{RRR}^{(0)}=f_{111},\qquad f_{RR\mathcal{G}}^{(0)}=f_{112},\\
f_{R\mathcal{G}\mathcal{G}}^{(0)}=f_{122},\qquad f_{\mathcal{G}\mathcal{G}\mathcal{G}}^{(0)}=f_{222}.
\end{array}\right.\tag{28}
\end{equation}
\subsection{Expanding the vacuum field equation and its trace equation in powers of $\overline{v}^2$}
In order to obtain general vacuum solutions up to $O(4)$ order, let us use the quantities in (\ref{10})--(\ref{28}) to expand the field equation and its trace equation, and then by (\ref{6}) and (\ref{7}), we obtain
\begin{equation}\label{29}
\left\{\begin{array}{ll}
H^{(0)}_{\mu\nu}=0,&H^{(0)}=0,\\
H^{(2)}_{\mu\nu}=0,&H^{(2)}=0,\\
H^{(3)}_{\mu\nu}=0,&H^{(3)}=0,\\
H^{(4)}_{\mu\nu}=0,&H^{(4)}=0
\end{array}\right.\tag{29}
\end{equation}
with
\begin{equation}\label{30}
\left\{\begin{array}{l}
H^{(0)}_{\mu\nu}=-\frac{g^{(0)}_{\mu\nu}}{2}f_{0},\\
H^{(0)}=-2f_{0},
\end{array}\right.\tag{30}
\end{equation}
\begin{equation}\label{31}
\left\{\begin{array}{l}
H^{(2)}_{tt}=-\frac{f_{1}}{2}R^{(2)}-\frac{f_{0}}{2}g^{(2)}_{tt}+f_{1}R^{(2)}_{tt}-f_{11}R^{(2)}_{,rr}-\frac{2f_{11}}{r}R^{(2)}_{,r},\\
H^{(2)}_{rr}=\frac{f_{1}}{2}R^{(2)}-\frac{f_{0}}{2}g^{(2)}_{rr}+f_{1}R^{(2)}_{rr}+\frac{2f_{11}}{r}R^{(2)}_{,r},\\
H^{(2)}_{\theta\theta}=\frac{f_{1}r^2}{2}R^{(2)}+f_{1}R^{(2)}_{\theta\theta}+f_{11}r^2R^{(2)}_{,rr}+f_{11}r R^{(2)}_{,r},\\
H^{(2)}_{\varphi\varphi}=\sin^2\theta H^{(2)}_{\theta\theta},\\
H^{(2)}=-f_{1}R^{(2)}-3f_{11}R^{(2)}_{,rr}-\frac{6f_{11}}{r}R^{(2)}_{,r},
\end{array}\right.\tag{31}
\end{equation}
\begin{equation}\label{32}
\left\{\begin{array}{l}
H^{(3)}_{tr}=f_{1}R^{(3)}_{tr}-f_{11}R^{(2)}_{,tr},\\
H^{(3)}=0,
\end{array}\right.\tag{32}
\end{equation}
and
\begin{widetext}
\begin{equation}\label{33}
\left\{\begin{array}{l}
H^{(4)}_{tt}=-\frac{f_{11}}{4}R^{(2)2}-\frac{f_{1}}{2}g^{(2)}_{tt}R^{(2)}-\frac{f_{1}}{2}R^{(4)}+f_{11}R^{(2)}R^{(2)}_{tt}+f_{1}R^{(4)}_{tt}
-\frac{2f_{12}}{r}\mathcal{G}^{(4)}_{,r}-f_{12}\mathcal{G}^{(4)}_{,rr}-\frac{4f_{12}}{r}R^{(2)}R^{(2)}_{,r}\\
\qquad\quad\ -\frac{2f_{111}}{r}R^{(2)}R^{(2)}_{,r}-\frac{2f_{11}}{r}g^{(2)}_{rr}R^{(2)}_{,r}
-\frac{2f_{11}}{r}g^{(2)}_{tt}R^{(2)}_{,r}+\frac{8f_{12}}{r}R^{(2)}_{,r}R^{(2)}_{tt}
-\frac{8f_{12}}{r^3}R^{(2)}_{,r}R^{(2)}_{\theta\theta}-f_{111}R^{(2)2}_{,r}\\
\qquad\quad\ -2f_{12}R^{(2)}R^{(2)}_{,rr}-f_{111}R^{(2)}R^{(2)}_{,rr}-f_{11}g^{(2)}_{rr}R^{(2)}_{,rr}
-f_{11}g^{(2)}_{tt}R^{(2)}_{,rr}-4f_{12}R^{(2)}_{rr}R^{(2)}_{,rr}+4f_{12}R^{(2)}_{tt}R^{(2)}_{,rr}\\
\qquad\quad\ -\frac{2f_{11}}{r}R^{(4)}_{,r}-f_{11}R^{(4)}_{,rr}-\frac{f_{11}}{2}R^{(2)}_{,r}g^{(2)}_{rr,r}
-\frac{4f_{12}}{r^2}R^{(2)}_{,r}g^{(2)}_{tt,r}-2f_{12}R^{(2)}_{,rr}g^{(2)}_{tt,rr}-\frac{f_{0}}{2}g^{(4)}_{tt},\\
H^{(4)}_{rr}=\frac{f_{11}}{4}R^{(2)2}-\frac{f_{1}}{2}g^{(2)}_{rr}R^{(2)}+\frac{f_{1}}{2}R^{(4)}+f_{11}R^{(2)}R^{(2)}_{rr}+f_{1}R^{(4)}_{rr}
+\frac{2f_{12}}{r}\mathcal{G}^{(4)}_{,r}+\frac{4f_{12}}{r}R^{(2)}R^{(2)}_{,r}+\frac{2f_{111}}{r}R^{(2)}R^{(2)}_{,r}\\
\qquad\quad\ +\frac{8f_{12}}{r}R^{(2)}_{,r}R^{(2)}_{rr}+\frac{8f_{12}}{r^3}R^{(2)}_{,r}R^{(2)}_{\theta\theta}
-f_{11}R^{(2)}_{,tt}+\frac{2f_{11}}{r}R^{(4)}_{,r}+\frac{4f_{12}}{r^2}R^{(2)}_{,r}g^{(2)}_{rr,r}
+\frac{f_{11}}{2}R^{(2)}_{,r}g^{(2)}_{tt,r}-\frac{f_{0}}{2}g^{(4)}_{rr},\\
H^{(4)}_{\theta\theta}=\frac{f_{11}r^2}{4}R^{(2)2}+f_{11}R^{(2)}R^{(2)}_{\theta\theta}+\frac{f_{1}r^2}{2}R^{(4)}
+f_{1}R^{(4)}_{\theta\theta}+f_{12}r\mathcal{G}^{(4)}_{,r}+f_{12}r^2\mathcal{G}^{(4)}_{,rr}+2f_{12}rR^{(2)}R^{(2)}_{,r}\\
\qquad\quad\ +f_{111}rR^{(2)}R^{(2)}_{,r}+f_{11}rg^{(2)}_{rr}R^{(2)}_{,r}+\frac{4f_{12}}{r}R^{(2)}_{,r}g^{(2)}_{rr}
+\frac{8f_{12}}{r}R^{(2)}_{\theta\theta}R^{(2)}_{,r}+f_{111}r^2R^{(2)2}_{,r}+2f_{12}r^2R^{(2)}R^{(2)}_{,rr}\\
\qquad\quad\ +f_{111}r^2R^{(2)}R^{(2)}_{,rr}+f_{11}r^2g^{(2)}_{rr}R^{(2)}_{,rr}+4f_{12}r^2R^{(2)}_{rr}R^{(2)}_{,rr}
+4f_{12}R^{(2)}_{\theta\theta}R^{(2)}_{,rr}-f_{11}r^2R^{(2)}_{,tt}+f_{11}rR^{(4)}_{,r}\\
\qquad\quad\ +f_{11}r^2R^{(4)}_{,rr}+\frac{f_{11}r^2}{2}R^{(2)}_{,r}g^{(2)}_{rr,r}+2f_{12}rR^{(2)}_{,rr}g^{(2)}_{rr,r}
+\frac{f_{11}r^2}{2}R^{(2)}_{,r}g^{(2)}_{tt,r},\\
H^{(4)}_{\varphi\varphi}=\sin^2\theta H^{(4)}_{\theta\theta},\\
H^{(4)}=-f_{1}R^{(4)}-\frac{6f_{12}}{r}\mathcal{G}^{(4)}_{,r}-3f_{12}\mathcal{G}^{(4)}_{,rr}-\frac{4f_{12}}{r}R^{(2)}R^{(2)}_{,r}
-\frac{6f_{111}}{r}R^{(2)}R^{(2)}_{,r}-\frac{6f_{11}}{r}g^{(2)}_{rr}R^{(2)}_{,r}
-\frac{8f_{12}}{r^3}R^{(2)}_{\theta\theta}R^{(2)}_{,r}\\
\qquad\quad\ -3f_{111}R^{(2)2}_{,r}-2f_{12}R^{(2)}R^{(2)}_{,rr}-3f_{111}R^{(2)}R^{(2)}_{,rr}
-3f_{11}g^{(2)}_{rr}R^{(2)}_{,rr}-4f_{12}R^{(2)}_{rr}R^{(2)}_{,rr}+3f_{11}R^{(2)}_{,tt}\\
\qquad\quad\ -\frac{6f_{11}}{r}R^{(4)}_{,r}-3f_{11}R^{(4)}_{,rr}-\frac{3f_{11}}{2}R^{(2)}_{,r}g^{(2)}_{rr,r}-\frac{3f_{11}}{2}R^{(2)}_{,r}g^{(2)}_{tt,r}.
\end{array}\right.\tag{33}
\end{equation}
\end{widetext}
Moreover by (\ref{20}), (\ref{31}), and (\ref{33}), there are
\begin{equation}\label{34}
\left\{\begin{array}{l}
H^{(2)}=H^{(2)}_{tt}-H^{(2)}_{rr}-\frac{2}{r^2}H^{(2)}_{\theta\theta}+\frac{f_{0}}{2}(g^{(2)}_{tt}-g^{(2)}_{rr}),\\
H^{(4)}=H^{(4)}_{tt}-H^{(4)}_{rr}-\frac{2}{r^2}H^{(4)}_{\theta\theta}-g^{(2)}_{tt}H^{(2)}_{tt}-g^{(2)}_{rr}H^{(2)}_{rr}\\
\qquad\quad\ +\frac{f_{0}}{2}(g^{(4)}_{tt}-g^{(4)}_{rr})-\frac{f_{0}}{2}(g^{(2)2}_{tt}+g^{(2)2}_{rr}).
\end{array}\right.\tag{34}
\end{equation}
\subsection{General spherically symmetric vacuum solutions for $f(R,\mathcal{G})$ gravity at $O(0)$, $O(2)$, and $O(3)$ order}
By (\ref{29}) and (\ref{30}), we know that the solution of $O(0)$ order equations is
\begin{equation}\label{35}
f_{0}=0,\tag{35}
\end{equation}
and then by (\ref{25}), this result shows that the cosmological constant contribution has to be zero whatever is the $f(R,\mathcal{G})$ gravity theory. The asymptotic cosmology is the de Sitter spacetime~\cite{9}, and then we can assume that it has a small positive constant curvature. By (\ref{5}) and (\ref{7}), we know that like $f(R)$ gravity~\cite{30}, the vacuum trace equation admits that $f(R,\mathcal{G})$ gravities has the de Sitter spacetime even when the cosmological constant contribution is zero. Therefore, our above assumption is compatible with (\ref{35}).

Now we start to deal with the differential equations at $O(2)$ order. By (\ref{20}), (\ref{29}), (\ref{31}), and (\ref{35}), the differential equations at $O(2)$ order are
\begin{equation}\label{36}
\left\{\begin{array}{l}
-\frac{f_{1}}{2}R^{(2)}+f_{1}R^{(2)}_{tt}-f_{11}R^{(2)}_{,rr}-\frac{2f_{11}}{r}R^{(2)}_{,r}=0,\\
\frac{f_{1}}{2}R^{(2)}+f_{1}R^{(2)}_{rr}+\frac{2f_{11}}{r}R^{(2)}_{,r}=0,\\
\frac{f_{1}r^2}{2}R^{(2)}+f_{1}R^{(2)}_{\theta\theta}+f_{11}r^2R^{(2)}_{,rr}+f_{11}r R^{(2)}_{,r}=0,\\
-f_{1}R^{(2)}-3f_{11}R^{(2)}_{,rr}-\frac{6f_{11}}{r}R^{(2)}_{,r}=0,\\
R^{(2)}=g^{(2)}_{tt,rr}+\frac{2}{r^2}g^{(2)}_{rr}+\frac{2}{r}g^{(2)}_{tt,r}+\frac{2}{r}g^{(2)}_{rr,r}.
\end{array}\right.\tag{36}
\end{equation}
Moreover, from the first equality of (\ref{34}) we know that there are only four independent differential equations in (\ref{36}). Equations (\ref{36}) are only related to $f_{1}$ and $f_{11}$, and then by (\ref{25}), we know that $\mathcal{G}$ has no influence on it and its solution. This implies that for $f(R,\mathcal{G})$ gravity and $f(R)$ gravity, both of their differential equations at $O(2)$ order are (\ref{36}). In addition, (\ref{36}) are only related to variable $r$, so both $f(R,\mathcal{G})$ gravity and $f(R)$ gravity have the same spatial behavior at $O(2)$ order. According the conclusions in Refs.~\cite{21,22}, these two behaviors are the Yukawa-like behavior and the oscillating-like behavior respectively.

Define parameter~\cite{23}
\begin{equation}\label{37}
m=\sqrt{\left|\frac{f_{1}}{3f_{11}}\right|}\tag{37}
\end{equation}
whose dimension is length$^{-1}$. We easily know that $m$ is also $m_{c}$ (Eq.~(26)) in Ref.~\cite{9} under the assumption $f(R=0)=0$. If $\text{sign}(f_{1})=-\text{sign}(f_{11})$, the solution of (\ref{36}) which corresponds to the Yukawa-like behavior is
\begin{equation}\label{38}
\left\{\begin{array}{l}
g^{(2)}_{tt}(t,r)=-\frac{C^{(2)}_{1}}{r}+g^{(2)}_{1}(t)\frac{e^{-mr}}{3m^2r},\\
g^{(2)}_{rr}(t,r)=-\frac{C^{(2)}_{1}}{r}-g^{(2)}_{1}(t)\frac{(mr+1)e^{-mr}}{3m^2r},\\
R^{(2)}(t,r)=g^{(2)}_{1}(t)\frac{e^{-mr}}{r}.
\end{array}\right.\tag{38}
\end{equation}
We know that it is asymptotic de Sitter spacetime with a small constant curvature~\cite{9}. In addition, if $\text{sign}(f_{1})=\text{sign}(f_{11})$, the solution of (\ref{36}) which corresponds to the oscillating-like behavior is
\begin{equation}\label{39}
\left\{\begin{array}{l}
g^{(2)}_{tt}(t,r)=-\frac{C^{(2)}_{2}}{r}-g^{(2)}_{2}(t)\frac{e^{-imr}}{3m^2r}+ig^{(2)}_{3}(t)\frac{e^{imr}}{6m^3r},\\
g^{(2)}_{rr}(t,r)=-\frac{C^{(2)}_{2}}{r}+g^{(2)}_{2}(t)\frac{(imr+1)e^{-imr}}{3m^2r}\\
\qquad\qquad\qquad\quad\ \ +ig^{(2)}_{3}(t)\frac{(imr-1)e^{imr}}{6m^3r},\\
R^{(2)}(t,r)=g^{(2)}_{2}(t)\frac{e^{-imr}}{r}-ig^{(2)}_{3}(t)\frac{e^{imr}}{2mr}.
\end{array}\right.\tag{39}
\end{equation}
Because of $g^{(2)}_{rr}$, we know that it is not asymptotic de Sitter spacetime with a small constant curvature in general. Both $C^{(2)}_{1}$ and $C^{(2)}_{2}$ are constants, and both of their dimensions are  length$^{1}$. All of $g^{(2)}_{1}(t)$, $g^{(2)}_{2}(t)$, and $g^{(2)}_{3}(t)$ are the function of time $t$, and the dimensions of $g^{(2)}_{1}(t)$, $g^{(2)}_{2}(t)$, and $g^{(2)}_{3}(t)$ are respectively length$^{-1}$, length$^{-1}$, and length$^{-2}$. Both (\ref{38}) and (\ref{39}) are time-dependent, and the time-dependent evolution depends on the the order of perturbations. So like $f(R)$ gravity
~\cite{21}, the Birkhoff theorem is no longer a general result for $f(R,\mathcal{G})$ gravity.

If the gravitational field is generated by a ball-like source described by (\ref{23}), when $f\rightarrow R$, the solution of (\ref{36}) should recover the perturbed version of standard Schwarzschild solution at $O(2)$ order, namely
\begin{equation}\label{40}
g^{(2)}_{tt}(t,r)=-\frac{r_{g}}{r},\quad g^{(2)}_{rr}(t,r)=-\frac{r_{g}}{r},\quad R^{(2)}(t,r)=0,\tag{40}
\end{equation}
where $r_{g}=2GM$. Then, we can easily verify that $C^{(2)}_{1}=r_{g}$ in (\ref{38}). But for (\ref{39}), when $f\rightarrow R$, we do not know whether it can recover (\ref{40}), as we can not confirm that whether there are $m^{n}$ in $g^{(2)}_{2}(t)$ and $g^{(2)}_{3}(t)$. So, we still need to consider the asymptotic behavior of (\ref{39}).
By increasing the distance $r$ from the ball-like source, the gravitational field (\ref{39}) should come close to the one in GR, namely (\ref{40}). By the first equality in (\ref{39}), there is $C^{(2)}_{2}=r_{g}$.

Since $\Phi=\frac{1}{2}g^{(2)}_{tt}$, where $\Phi$ is the gravitational potential, (\ref{38}) and (\ref{39}) can
provide two kinds of corrected gravitational potentials with respect to the Newtonian one $-\frac{GM}{r}$, namely
\begin{equation}\label{41}
\left\{\begin{array}{l}
\Phi_{\mathrm{Yukawa}}=-\frac{GM}{r}+g^{(2)}_{1}(t)\frac{e^{-mr}}{6m^2r},\\
\Phi_{\mathrm{oscillating}}=-\frac{GM}{r}-g^{(2)}_{2}(t)\frac{e^{-imr}}{6m^2r}+ig^{(2)}_{3}(t)\frac{e^{imr}}{12m^3r}.
\end{array}\right.\tag{41}
\end{equation}

In Ref.~\cite{25}, the Newtonian limit of the most general fourth-order theory of gravity, namely $F(X,Y,Z)$ gravity, has been studied, with $X=R$, $Y=R_{\mu\nu}R^{\mu\nu}$, and $Z=R_{\mu\nu\sigma\rho}R^{\mu\nu\sigma\rho}$. By its conclusions, for the gravitational field generated by a ball-like source described by (\ref{23}), the static Yukawa-like behavior has two characteristic lengths $m_{1}^{-1}$ and $m_{2}^{-1}$, where $m_{1}$ and $m_{2}$ are defined as (\ref{1}) and (\ref{2}). Because both $f(R,\mathcal{G})$ gravity and $f(R)$ gravity have the same spatial behavior at $O(2)$ order, if $g^{(2)}_{1}(t)=-r_{g}m^{2}$~\cite{23}, we obtain the static Yukawa-like behavior for $f(R,\mathcal{G})$ gravity at $O(2)$ order, namely
\begin{equation}\label{42}
\left\{\begin{array}{l}
g^{(2)}_{tt}(r)=-\frac{r_{g}}{r}-r_{g}\frac{e^{-mr}}{3r},\\
g^{(2)}_{rr}(r)=-\frac{r_{g}}{r}+r_{g}\frac{(mr+1)e^{-mr}}{3r},\\
R^{(2)}(r)=-r_{g}m^{2}\frac{e^{-mr}}{r}.
\end{array}\right.\tag{42}
\end{equation}
Obviously, there is only one characteristic lengths $m^{-1}$ in (\ref{42}). Now we will prove that if $F(X,Y,Z)=f(R,\mathcal{G})$, there are
\begin{equation}\label{43}
m_{1}=m,\qquad m_{2}=\infty.\tag{43}
\end{equation}
In fact, by (\ref{3}) and (\ref{26}), there are equalities
\begin{align*}
&F_{X}(0,0,0)=f_{R}(0,0)=f_{1},\\
&F_{XX}(0,0,0)=f_{RR}(0,0)+2f_{\mathcal{G}}(0,0)=f_{11}+2f_{2},\\
&F_{Y}(0,0,0)=-4f_{\mathcal{G}}(0,0)=-4f_{2},\\
&F_{Z}(0,0,0)=f_{\mathcal{G}}(0,0)=f_{2},
\end{align*}
and then by (\ref{1}), (\ref{2}), and (\ref{37}), we know that (\ref{43}) hold. Applying (\ref{43}) to (19) in Ref.~\cite{25}, we know that our result (\ref{42}) is compatible with it by $g^{(2)}_{tt}$ and $R^{(2)}$ (we have set $f_{1}=1$ for simplicity). Because the system of coordinates in Ref.~\cite{25} is different from ours, $g^{(2)}_{rr}$ in Ref.~\cite{25} is different from our result in (\ref{42}). From (\ref{43}), we know that because of the presence of the GB invariant, the characteristic length $m_{2}^{-1}$ disappears, and the characteristic length $m_{1}^{-1}$ has nothing to do with the quadratic curvature invariants $X^{2}$, $Y$, and $Z$.
\begin{figure}[t!]
\centering
\includegraphics[scale=0.67]{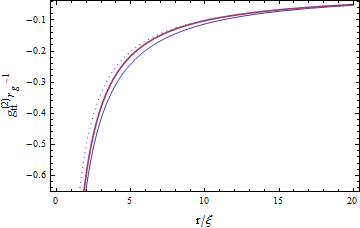}
\caption{\label{fig:1} Plot of the spatial behaviors of $g^{(2)}_{tt}$ for $f(R,\mathcal{G})$ gravity and GR in the Yukawa-like case. The dotted curve is the behavior of GR; the thin solid curve and the thick solid curve are the behaviors of $f(R,\mathcal{G})$ gravity with $m$=0.1,~0.3.}
\end{figure}
\begin{figure}[b!]
\centering
\includegraphics[scale=0.67]{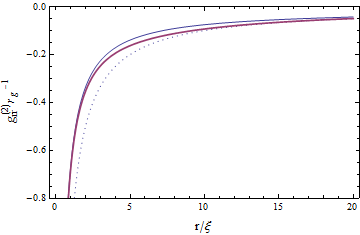}
\caption{\label{fig:2} Plot of the spatial behaviors of $g^{(2)}_{rr}$ for $f(R,\mathcal{G})$ gravity and GR in the Yukawa-like case. The dotted curve is the behavior of GR; the thin solid curve and the thick solid curve are the behaviors of $f(R,\mathcal{G})$ gravity with $m$=0.1,~0.3.}
\end{figure}
\begin{figure}[t!]
\centering
\includegraphics[scale=0.67]{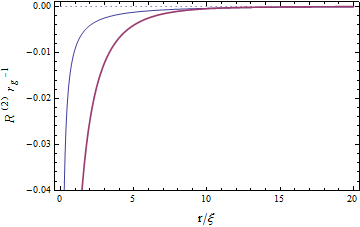}
\caption{\label{fig:3} Plot of the spatial behaviors of $R^{(2)}$ for $f(R,\mathcal{G})$ gravity and GR in the Yukawa-like case. The dotted line is the behavior of GR; the thin solid curve and the thick solid curve are the behaviors of $f(R,\mathcal{G})$ gravity with $m$=0.1,~0.3.}
\end{figure}
\begin{figure}[b!]
\centering
\includegraphics[scale=0.67]{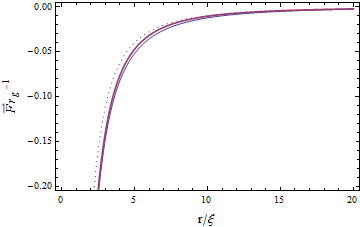}
\caption{\label{fig:4} Plot of the spatial behaviors of gravitational force $\vec{F}$ for $f(R,\mathcal{G})$ gravity and GR in the Yukawa-like case. The dotted curve is the behavior of GR; the thin solid curve and the thick solid curve are the behaviors of $f(R,\mathcal{G})$ gravity with $m$=0.1,~0.3.}
\end{figure}

In order to compare (\ref{40}) with (\ref{42}), the spatial behaviors of $g^{(2)}_{tt}$, $g^{(2)}_{rr}$, and $R^{(2)}$
for $f(R,\mathcal{G})$ gravity and GR in the Yukawa-like case are shown in Figs.~\ref{fig:1}--\ref{fig:3}. The gravitational forces induced by $f(R,\mathcal{G})$ gravity and GR in the Yukawa-like case are shown in Fig.~\ref{fig:4}. From Fig.~\ref{fig:1}, we find that the gravitational potentials have the divergency at $r=0$, but this is different from the corresponding conclusion about $F(X,Y,Z)$ gravity in Ref.~\cite{25}.

Compared with (\ref{39}), the real valued behavior in it interests us more. In (\ref{39}),
\begin{equation}\label{44}
\left\{\begin{array}{l}
g^{(2)}_{tt}(t,r)=-\frac{C^{(2)}_{2}}{r}-\frac{R^{(2)}(t,r)}{3m^2},\\
g^{(2)}_{rr}(t,r)=-\frac{C^{(2)}_{2}}{r}-\frac{rR^{(2)}_{,r}(t,r)}{3m^2}\\
\end{array}\right.\tag{44}
\end{equation}
hold, so the real valued oscillating-like behavior is
\begin{equation}\label{45}
\left\{\begin{array}{l}
g^{(2)}_{tt}(t,r)=-\frac{C^{(2)}_{2}}{r}-g^{(2)}_{4}(t)\frac{\cos(mr)}{3m^2r}-g^{(2)}_{5}(t)\frac{\sin(mr)}{3m^2r},\\
g^{(2)}_{rr}(t,r)=-\frac{C^{(2)}_{2}}{r}+(g^{(2)}_{4}(t)-mrg^{(2)}_{5}(t))\frac{\cos(mr)}{3m^2r}\\
\qquad\qquad\quad\ \ +(mrg^{(2)}_{4}(t)+g^{(2)}_{5}(t))\frac{\sin(mr)}{3m^2r},\\
R^{(2)}(t,r)=g^{(2)}_{4}(t)\frac{\cos(mr)}{r}+g^{(2)}_{5}(t)\frac{\sin(mr)}{r},
\end{array}\right.\tag{45}
\end{equation}
where both $g^{(2)}_{4}(t)$ and $g^{(2)}_{5}(t)$ are the function of time $t$, and both of their dimensions are length$^{-1}$.

Because both $f(R,\mathcal{G})$ gravity and $f(R)$ gravity have the same spatial behavior at $O(2)$ order, for the gravitational field generated by a ball-like source described by (\ref{23}), if $g^{(2)}_{4}(t)=g^{(2)}_{5}(t)=-r_{g}m^{2}$~\cite{23}, we obtain the real valued and static oscillating-like behavior for $f(R,\mathcal{G})$ gravity at $O(2)$ order, namely
\begin{equation}\label{46}
\left\{\begin{array}{l}
g^{(2)}_{tt}(r)=-\frac{r_{g}}{r}+r_{g}\left(\frac{\cos(mr)}{3r}+\frac{\sin(mr)}{3r}\right),\\
g^{(2)}_{rr}(r)=-\frac{r_{g}}{r}-r_{g}\left(\frac{\cos(mr)}{3r}+\frac{\sin(mr)}{3r}\right)\\
\qquad\qquad\quad+mr_{g}\left(\frac{\cos(mr)}{3}-\frac{\sin(mr)}{3}\right),\\
R^{(2)}(r)=-r_{g}m^{2}\left(\frac{\cos(mr)}{r}+\frac{\sin(mr)}{r}\right).
\end{array}\right.\tag{46}
\end{equation}
\begin{figure}[t!]
\centering
\includegraphics[scale=0.67]{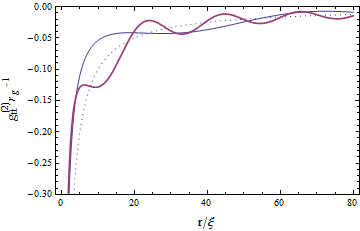}
\caption{\label{fig:5} Plot of the spatial behaviors of $g^{(2)}_{tt}$ for $f(R,\mathcal{G})$ gravity and GR in the oscillating-like case. The dotted curve is the behavior of GR; the thin solid curve and the thick solid curve are the behaviors of $f(R,\mathcal{G})$ gravity with $m$=0.1,~0.3.}
\end{figure}
\begin{figure}[b!]
\centering
\includegraphics[scale=0.67]{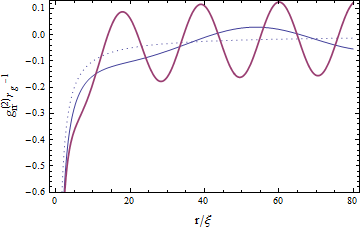}
\caption{\label{fig:6} Plot of the spatial behaviors of $g^{(2)}_{rr}$ for $f(R,\mathcal{G})$ gravity and GR in the oscillating-like case. The dotted curve is the behavior of GR; the thin solid curve and the thick solid curve are the behaviors of $f(R,\mathcal{G})$ gravity with $m$=0.1,~0.3.}
\end{figure}
\begin{figure}[t!]
\centering
\includegraphics[scale=0.67]{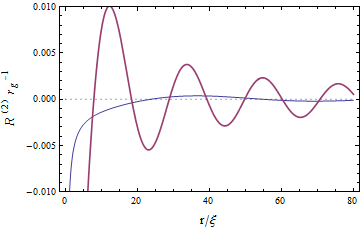}
\caption{\label{fig:7} Plot of the spatial behaviors of $R^{(2)}$ for $f(R,\mathcal{G})$ gravity and GR in the oscillating-like case. The dotted line is the behavior of GR; the thin solid curve and the thick solid curve are the behaviors of $f(R,\mathcal{G})$ gravity with $m$=0.1,~0.3.}
\end{figure}
\begin{figure}[b!]
\centering
\includegraphics[scale=0.67]{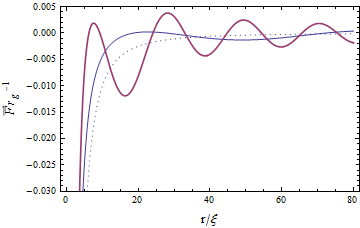}
\caption{\label{fig:8} Plot of the spatial behaviors of gravitational force $\vec{F}$ for $f(R,\mathcal{G})$ gravity and GR in the oscillating-like case. The dotted curve is the behavior of GR; the thin solid curve and the thick solid curve are the behaviors of $f(R,\mathcal{G})$ gravity with $m$=0.1,~0.3.}
\end{figure}
It is easy to know that (\ref{46}) can not recover the standard Schwarzschild solution (\ref{40}) at $O(2)$ order  when $f\rightarrow R$. We need to note that (\ref{46}) does not come as a specific consequence of using the Schwarzschild coordinates, as in Ref.~\cite{23}, there is corresponding result about $f(R)$ gravity with the isotropic coordinates. Therefore (\ref{46}) is a natural result of $f(R,\mathcal{G}))$ gravity and $f(R)$ gravity.

In order to compare (\ref{40}) with (\ref{46}), the spatial behaviors of $g^{(2)}_{tt}$, $g^{(2)}_{rr}$, and $R^{(2)}$
for $f(R,\mathcal{G})$ gravity and GR in the oscillating-like case are shown in Figs.~\ref{fig:5}--\ref{fig:7}. The gravitational forces induced by $f(R,\mathcal{G})$ gravity and GR are shown in Fig.~\ref{fig:8}.

In order to parameterize the deviation with respect to GR for $f(R,\mathcal{G})$ gravity, we show its Eddington-parameters by the static behaviors (\ref{42}) and (\ref{46}). The Eddington-parameters that are associated with the Schwarzschild coordinates are defined by~\cite{29}
\begin{align*}
ds^{2}\simeq& [1-\frac{r_{g}}{r}+\frac{\beta-\gamma}{2}(\frac{r_{g}}{r})^{2}+\cdots]dt^{2}\\&-[1+\gamma\frac{r_{g}}{r}+\cdots]dr^{2}-r^{2}d\Omega^{2}.
\end{align*}
By Ref.~\cite{29}, $\gamma$ measures the amount of curvature of space generated by a body of mass $M$ at radius $r$, and $\beta$ measures the amount of non-linearity ($\sim (r_{g}/r)^{2}$) in the $g_{tt}$ component of the metric, but the latter is valid only in the standard post-Newtonian gauge. Therefore, here we only discuss $\gamma$, which can be obtained within $O(2)$ order. By (\ref{11}), (\ref{42}), and (\ref{46}), the Eddingtion parameters $\gamma$ of $f(R,\mathcal{G})$ gravity in the Yukawa-like case and in the oscillating-like case are
\begin{equation*}
\left\{\begin{array}{l}
\gamma_{\mathrm{Yukawa}}=1-\frac{mr+1}{3}e^{-mr},\\
\gamma_{\mathrm{oscillating}}=1+\left(\frac{\cos(mr)}{3}+\frac{\sin(mr)}{3}\right)\\
\qquad\qquad\quad\quad-mr\left(\frac{\cos(mr)}{3}-\frac{\sin(mr)}{3}\right).
\end{array}\right.
\end{equation*}
Above expressions show that $\gamma$ of $f(R,\mathcal{G})$ gravity is the function of the point but not a constant, and such result is similar to the corresponding conclusion about $F(X,Y,Z)$ gravity in Ref.~\cite{25}. Moreover, these expressions show that when $f\rightarrow R$, $\gamma_{\mathrm{Yukawa}}$ can recover the standard Schwarzschild result $\gamma=1$, but $\gamma_{\mathrm{oscillating}}$ can not. This is also compatible with the corresponding conclusions about (\ref{42}) and (\ref{46}).

Now we start to deal with the differential equation at $O(3)$ order. By (\ref{20}), (\ref{29}), (\ref{32}), and (\ref{35}), the differential equation at $O(3)$ order is
\begin{equation}\label{47}
f_{1}R^{(3)}_{tr}-f_{11}R^{(2)}_{,tr}=0.\tag{47}
\end{equation}
Equation (\ref{47}) is only related to $f_{1}$ and $f_{11}$, so by (\ref{25}), $\mathcal{G}$ has no influence on it and its solution, and then we know that both $f(R,\mathcal{G})$ gravity and $f(R)$ gravity have the same solution at $O(3)$ order. But unlike equations (\ref{36}), we know that (\ref{47}) is a differential equation about time $t$, so perhaps by it, we may fix $g^{(2)}_{1}(t)$, $g^{(2)}_{2}(t)$, or $g^{(2)}_{3}(t)$ in (\ref{38}) and (\ref{39}). In fact, this is impossible. By (\ref{20}), (\ref{47}) becomes
\begin{displaymath}
-\frac{f_{1}}{r}g^{(2)}_{rr,t}-f_{11}R^{(2)}_{,tr}=0,
\end{displaymath}
and then by (\ref{37}), we can verify that (\ref{38}) and (\ref{39}) satisfy above differential equation. In fact, the differential equations at $O(4)$ order still can not fix $g^{(2)}_{1}(t)$, $g^{(2)}_{2}(t)$, and $g^{(2)}_{3}(t)$, and this implies that in order to probe the time-dependent evolution about two behaviors at $O(2)$ order, we need to develop the WFSM limit of $f(R,\mathcal{G})$ gravity up to more orders.
\subsection{General spherically symmetric vacuum solution for $f(R,\mathcal{G})$ gravity at $O(4)$ order}
The solution of the $O(4)$ order differential equations is the correction to the result at $O(2)$ order. According to  the solution, we can know about the deviations with respect to GR for $f(R,\mathcal{G})$ gravity on a more precise level and the difference between $f(R,\mathcal{G})$ gravity and $f(R)$ gravity. By (\ref{20}), (\ref{29}), (\ref{33}) and (\ref{35}), the  differential equations at $O(4)$ order are
\begin{widetext}
\begin{equation}\label{48}
\left\{\begin{array}{l}
-\frac{f_{11}}{4}R^{(2)2}-\frac{f_{1}}{2}g^{(2)}_{tt}R^{(2)}-\frac{f_{1}}{2}R^{(4)}+f_{11}R^{(2)}R^{(2)}_{tt}+f_{1}R^{(4)}_{tt}
-\frac{2f_{12}}{r}\mathcal{G}^{(4)}_{,r}-f_{12}\mathcal{G}^{(4)}_{,rr}-\frac{4f_{12}}{r}R^{(2)}R^{(2)}_{,r}\\
-\frac{2f_{111}}{r}R^{(2)}R^{(2)}_{,r}-\frac{2f_{11}}{r}g^{(2)}_{rr}R^{(2)}_{,r}
-\frac{2f_{11}}{r}g^{(2)}_{tt}R^{(2)}_{,r}+\frac{8f_{12}}{r}R^{(2)}_{,r}R^{(2)}_{tt}
-\frac{8f_{12}}{r^3}R^{(2)}_{,r}R^{(2)}_{\theta\theta}-f_{111}R^{(2)2}_{,r}\\
-2f_{12}R^{(2)}R^{(2)}_{,rr}-f_{111}R^{(2)}R^{(2)}_{,rr}-f_{11}g^{(2)}_{rr}R^{(2)}_{,rr}
-f_{11}g^{(2)}_{tt}R^{(2)}_{,rr}-4f_{12}R^{(2)}_{rr}R^{(2)}_{,rr}+4f_{12}R^{(2)}_{tt}R^{(2)}_{,rr}\\
-\frac{2f_{11}}{r}R^{(4)}_{,r}-f_{11}R^{(4)}_{,rr}-\frac{f_{11}}{2}R^{(2)}_{,r}g^{(2)}_{rr,r}
-\frac{4f_{12}}{r^2}R^{(2)}_{,r}g^{(2)}_{tt,r}-2f_{12}R^{(2)}_{,rr}g^{(2)}_{tt,rr}=0,\\
\frac{f_{11}}{4}R^{(2)2}-\frac{f_{1}}{2}g^{(2)}_{rr}R^{(2)}+\frac{f_{1}}{2}R^{(4)}+f_{11}R^{(2)}R^{(2)}_{rr}+f_{1}R^{(4)}_{rr}
+\frac{2f_{12}}{r}\mathcal{G}^{(4)}_{,r}+\frac{4f_{12}}{r}R^{(2)}R^{(2)}_{,r}+\frac{2f_{111}}{r}R^{(2)}R^{(2)}_{,r}\\
+\frac{8f_{12}}{r}R^{(2)}_{,r}R^{(2)}_{rr}+\frac{8f_{12}}{r^3}R^{(2)}_{,r}R^{(2)}_{\theta\theta}
-f_{11}R^{(2)}_{,tt}+\frac{2f_{11}}{r}R^{(4)}_{,r}+\frac{4f_{12}}{r^2}R^{(2)}_{,r}g^{(2)}_{rr,r}
+\frac{f_{11}}{2}R^{(2)}_{,r}g^{(2)}_{tt,r}=0,\\
\frac{f_{11}r^2}{4}R^{(2)2}+f_{11}R^{(2)}R^{(2)}_{\theta\theta}+\frac{f_{1}r^{2}}{2}R^{(4)}
+f_{1}R^{(4)}_{\theta\theta}+f_{12}r\mathcal{G}^{(4)}_{,r}+f_{12}r^2\mathcal{G}^{(4)}_{,rr}
+2f_{12}rR^{(2)}R^{(2)}_{,r}+f_{111}rR^{(2)}R^{(2)}_{,r}\\
+f_{11}rg^{(2)}_{rr}R^{(2)}_{,r}+\frac{4f_{12}}{r}R^{(2)}_{,r}g^{(2)}_{rr}
+\frac{8f_{12}}{r}R^{(2)}_{\theta\theta}R^{(2)}_{,r}+f_{111}r^2R^{(2)2}_{,r}+2f_{12}r^2R^{(2)}R^{(2)}_{,rr}
+f_{111}r^2R^{(2)}R^{(2)}_{,rr}\\
+f_{11}r^2g^{(2)}_{rr}R^{(2)}_{,rr}+4f_{12}r^2R^{(2)}_{rr}R^{(2)}_{,rr}
+4f_{12}R^{(2)}_{\theta\theta}R^{(2)}_{,rr}-f_{11}r^2R^{(2)}_{,tt}+f_{11}rR^{(4)}_{,r}+f_{11}r^2R^{(4)}_{,rr}\\
+\frac{f_{11}r^2}{2}R^{(2)}_{,r}g^{(2)}_{rr,r}+2f_{12}rR^{(2)}_{,rr}g^{(2)}_{rr,r}
+\frac{f_{11}r^2}{2}R^{(2)}_{,r}g^{(2)}_{tt,r}=0,\\
-f_{1}R^{(4)}-\frac{6f_{12}}{r}\mathcal{G}^{(4)}_{,r}-3f_{12}\mathcal{G}^{(4)}_{,rr}-\frac{4f_{12}}{r}R^{(2)}R^{(2)}_{,r}
-\frac{6f_{111}}{r}R^{(2)}R^{(2)}_{,r}-\frac{6f_{11}}{r}g^{(2)}_{rr}R^{(2)}_{,r}
-\frac{8f_{12}}{r^3}R^{(2)}_{\theta\theta}R^{(2)}_{,r}\\
-3f_{111}R^{(2)2}_{,r}-2f_{12}R^{(2)}R^{(2)}_{,rr}-3f_{111}R^{(2)}R^{(2)}_{,rr}
-3f_{11}g^{(2)}_{rr}R^{(2)}_{,rr}-4f_{12}R^{(2)}_{rr}R^{(2)}_{,rr}+3f_{11}R^{(2)}_{,tt}\\ -\frac{6f_{11}}{r}R^{(4)}_{,r}-3f_{11}R^{(4)}_{,rr}-\frac{3f_{11}}{2}R^{(2)}_{,r}g^{(2)}_{rr,r}-\frac{3f_{11}}{2}R^{(2)}_{,r}g^{(2)}_{tt,r}=0,\\
R^{(4)}=g^{(4)}_{tt,rr}+\frac{2}{r^2}g^{(4)}_{rr}+\frac{2}{r}g^{(4)}_{tt,r}+\frac{2}{r}g^{(4)}_{rr,r}
-g^{(2)}_{tt}g^{(2)}_{tt,rr}+g^{(2)}_{rr}g^{(2)}_{tt,rr}+\frac{2}{r^2}g^{(2)2}_{rr}-\frac{2}{r}g^{(2)}_{tt}g^{(2)}_{tt,r}+\frac{2}{r}g^{(2)}_{rr}g^{(2)}_{tt,r}\\
+\frac{4}{r}g^{(2)}_{rr}g^{(2)}_{rr,r}+\frac{1}{2}g^{(2)}_{tt,r}g^{(2)}_{rr,r}-
\frac{1}{2}g^{(2)2}_{tt,r}+g^{(2)}_{rr,tt}.
\end{array}\right.\tag{48}
\end{equation}
\end{widetext}
Moreover, from the second equality of (\ref{34}), we know that there are only four independent differential equations in (\ref{48}). Equations (\ref{48}) are only related to $f_{1}$, $f_{11}$, $f_{12}$, and $f_{111}$, and then by (\ref{25}), we know that $\mathcal{G}$ is only associated with the term $R\mathcal{G}$ in the above Taylor expansion of $f$. Because $f_{12}=0$ in (\ref{25}) for $f(R)$ gravity, this implies that the solutions of (\ref{48}) for $f(R,\mathcal{G})$ gravity and $f(R)$ gravity are completely different.

After putting (\ref{38}) or (\ref{46}) into (\ref{48}), we know that (\ref{48}) are the differential equations related to $g^{(4)}_{tt}$, $g^{(4)}_{rr}$, and $R^{(4)}$, which are the correction to the results at $O(2)$ order. To obtain these three quantities, we only need three differential equations in (\ref{48}). Firstly, by the fourth differential equation in (\ref{48}), we can obtain $R^{(4)}$. Next, put it into the the first differential equation and the second differential equation in (\ref{48}), and then we can obtain $g^{(4)}_{tt}$ and $g^{(4)}_{rr}$. Finally, we put $g^{(4)}_{tt}$, $g^{(4)}_{rr}$, and $R^{(4)}$ above into the remaining two differential equations in (\ref{48}) to eliminate redundant quantities in their expressions. Thus, we obtain the corrections to (\ref{38}) and (\ref{46}). Because the correction to (\ref{46}) is so involuted that we do not want to present it. Here we only present the correction to (\ref{38}). Define parameters
\begin{equation}\label{49}
\alpha_{1}=\frac{f_{12}}{f_{1}},\qquad \alpha_{2}=\frac{f_{111}}{f_{1}},\tag{49}
\end{equation}
where both of their dimensions are length$^{4}$. In order to obtain the right result, we need to set the solution of (\ref{48}) to be de Sitter spacetime with a small constant curvature in the asymptotic region and can recover the perturbed version of standard Schwarzschild solution at $O(4)$ order for a gravitational field generated by a ball-like source described by (\ref{23}), namely
\begin{equation}\label{50}
g^{(4)}_{tt}(t,r)=0,\quad g^{(4)}_{rr}(t,r)=-\frac{r_{g}^{2}}{r^2},\quad R^{(4)}(t,r)=0.\tag{50}
\end{equation}
The correction to (\ref{38}) is
\begin{equation}\label{51}
\left\{\begin{array}{l}
g^{(4)}_{tt}(t,r)=\phi_{tt}(t,r)+\alpha_{1} m^{4}\phi_{tt1}(t,r)+\alpha_{2} m^{4}\phi_{tt2}(t,r),\\
g^{(4)}_{rr}(t,r)=\phi_{rr}(t,r)+\alpha_{1} m^{4}\phi_{rr1}(t,r)+\alpha_{2} m^{4}\phi_{rr2}(t,r),\\
R^{(4)}(t,r)=\phi(t,r)+\alpha_{1} m^{4}\phi_{1}(t,r)+\alpha_{2} m^{4}\phi_{2}(t,r),
\end{array}\right.\tag{51}
\end{equation}
where the above unknown functions are defined as
\begin{widetext}
\begin{equation}\label{52}
\left\{\begin{array}{l}
\phi_{tt}(t,r)=g^{(4)}_{1}(t)\frac{e^{-mr}}{3m^2r}-g^{(2)}_{1}(t)\frac{C^{(2)}_{1}e^{-mr}}{3m^2r^2}+g^{(2)''}_{1}(t)\frac{e^{-mr}}{6m^3}-g^{(2)''}_{1}(t)\frac{e^{-mr}}{12m^4r}+g^{(2)2}_{1}(t)\frac{e^{-2mr}}{9m^4r^2}\\
\phantom{\phi_{tt}(t,r)=}-g^{(2)2}_{1}(t)\frac{e^{-2mr}}{36m^3r}-g^{(2)}_{1}(t)\frac{C^{(2)}_{1}e^{-mr}\text{ln}(r)}{6mr}+g^{(2)}_{1}(t)\frac{C^{(2)}_{1}e^{mr}\text{Ei}(-2mr)}{6mr}+g^{(2)2}_{1}(t)\frac{e^{mr}\text{Ei}(-3mr)}{12m^3r}\\
\phantom{\phi_{tt}(t,r)=}-g^{(2)2}_{1}(t)\frac{\text{Ei}(-2mr)}{6m^2}-g^{(2)2}_{1}(t)\frac{e^{-mr}\text{Ei}(-mr)}{12m^3r},\\
\phi_{tt1}(t,r)=\frac{C^{(2)2}_{1}}{m^2r^4}+\frac{C^{(2)2}_{1}}{2r^2}-2g^{(2)}_{1}(t)\frac{C^{(2)}_{1}e^{-mr}}{m^4r^4}-2g^{(2)}_{1}(t)\frac{C^{(2)}_{1}e^{-mr}}{m^3r^3}+g^{(2)}_{1}(t)\frac{C^{(2)}_{1}e^{-mr}}{m^2r^2}-g^{(2)}_{1}(t)\frac{C^{(2)}_{1}e^{-mr}}{mr}\\
\phantom{\phi_{tt1}(t,r)=}-g^{(2)2}_{1}(t)\frac{e^{-2mr}}{3m^6r^4}-2g^{(2)2}_{1}(t)\frac{e^{-2mr}}{3m^5r^3}-g^{(2)2}_{1}(t)\frac{e^{-2mr}}{6m^4r^2}+\frac{C^{(2)2}_{1}me^{mr}\text{Ei}(-mr)}{4r}-\frac{C^{(2)2}_{1}me^{-mr}\text{Ei}(mr)}{4r}\\
\phantom{\phi_{tt1}(t,r)=}+g^{(2)2}_{1}(t)\frac{e^{mr}\text{Ei}(-3mr)}{4m^3r}-g^{(2)2}_{1}(t)\frac{e^{-mr}\text{Ei}(-mr)}{4m^3r}-g^{(2)}_{1}(t)C^{(2)}_{1}\text{Ei}(-mr),\\
\phi_{tt2}(t,r)=g^{(2)2}_{1}(t)\frac{e^{mr}\text{Ei}(-3mr)}{4m^3r}-g^{(2)2}_{1}(t)\frac{e^{-mr}\text{Ei}(-mr)}{4m^3r},\\
\phi_{rr}(t,r)=-\frac{C^{(2)2}_{1}}{r^2}-g^{(4)}_{1}(t)\frac{e^{-mr}}{3m}-g^{(4)}_{1}(t)\frac{e^{-mr}}{3m^2r}-5g^{(2)}_{1}(t)\frac{C^{(2)}_{1}e^{-mr}}{6m^2r^2}-g^{(2)}_{1}(t)\frac{C^{(2)}_{1}e^{-mr}}{2mr}+g^{(2)''}_{1}(t)\frac{e^{-mr}}{12m^3}\\
\phantom{\phi_{rr}(t,r)=}+g^{(2)''}_{1}(t)\frac{e^{-mr}}{12m^4r}+g^{(2)''}_{1}(t)\frac{re^{-mr}}{6m^2}-5g^{(2)2}_{1}(t)\frac{e^{-2mr}}{36m^2}-g^{(2)2}_{1}(t)\frac{e^{-2mr}}{18m^4r^2}-5g^{(2)2}_{1}(t)\frac{e^{-2mr}}{36m^3r}\\
\phantom{\phi_{rr}(t,r)=}+\frac{1}{6}g^{(2)}_{1}(t)C^{(2)}_{1}e^{-mr}\text{ln}(r)+g^{(2)}_{1}(t)\frac{C^{(2)}_{1}e^{-mr}\text{ln}(r)}{6mr}+\frac{1}{6}g^{(2)}_{1}(t)C^{(2)}_{1}e^{mr}\text{Ei}(-2mr)\\
\phantom{\phi_{rr}(t,r)=}-g^{(2)}_{1}(t)\frac{C^{(2)}_{1}e^{mr}\text{Ei}(-2mr)}{6mr}+g^{(2)2}_{1}(t)\frac{e^{mr}\text{Ei}(-3mr)}{12m^2}-g^{(2)2}_{1}(t)\frac{e^{mr}\text{Ei}(-3mr)}{12m^3r}+g^{(2)2}_{1}(t)\frac{e^{-mr}\text{Ei}(-mr)}{12m^2}\\
\phantom{\phi_{rr}(t,r)=}+g^{(2)2}_{1}(t)\frac{e^{-mr}\text{Ei}(-mr)}{12m^3r},\\
\phi_{rr1}(t,r)=-\frac{C^{(2)2}_{1}}{r^2}-\frac{4C^{(2)2}_{1}}{m^2r^4}-4g^{(2)}_{1}(t)\frac{C^{(2)}_{1}e^{-mr}}{m^4r^4}-4g^{(2)}_{1}(t)\frac{C^{(2)}_{1}e^{-mr}}{m^3r^3}+g^{(2)2}_{1}(t)\frac{e^{-2mr}}{3m^4r^2}+g^{(2)2}_{1}(t)\frac{e^{-2mr}}{3m^3r}\\
\phantom{\phi_{rr1}(t,r)=}+\frac{1}{4}C^{(2)2}_{1}m^2e^{mr}\text{Ei}(-mr)+\frac{1}{4}C^{(2)2}_{1}m^2e^{-mr}\text{Ei}(mr)-\frac{C^{(2)2}_{1}me^{mr}\text{Ei}(-mr)}{4r}+\frac{C^{(2)2}_{1}me^{-mr}\text{Ei}(mr)}{4r}\\
\phantom{\phi_{rr1}(t,r)=}+g^{(2)2}_{1}(t)\frac{e^{mr}\text{Ei}(-3mr)}{4m^2}-g^{(2)2}_{1}(t)\frac{e^{mr}\text{Ei}(-3mr)}{4m^3r}+g^{(2)2}_{1}(t)\frac{e^{-mr}\text{Ei}(-mr)}{4m^2}+g^{(2)2}_{1}(t)\frac{e^{-mr}\text{Ei}(-mr)}{4m^3r},\\
\phi_{rr2}(t,r)=g^{(2)2}_{1}(t)\frac{e^{mr}\text{Ei}(-3mr)}{4m^2}-g^{(2)2}_{1}(t)\frac{e^{mr}\text{Ei}(-3mr)}{4m^3r}+g^{(2)2}_{1}(t)\frac{e^{-mr}\text{Ei}(-mr)}{4m^2}+g^{(2)2}_{1}(t)\frac{e^{-mr}\text{Ei}(-mr)}{4m^3r},\\
\phi(t,r)=g^{(4)}_{1}(t)\frac{e^{-mr}}{r}+g^{(2)}_{1}(t)\frac{C^{(2)}_{1}e^{-mr}}{2r^2}-g^{(2)''}_{1}(t)\frac{e^{-mr}}{2m}-g^{(2)''}_{1}(t)\frac{e^{-mr}}{4m^2r}+g^{(2)2}_{1}(t)\frac{e^{-2mr}}{6mr}\\
\phantom{\phi(t,r)=}-g^{(2)}_{1}(t)\frac{C^{(2)}_{1}me^{-mr}\text{ln}(r)}{2r}+g^{(2)}_{1}(t)\frac{C^{(2)}_{1}me^{mr}\text{Ei}(-2mr)}{2r}+g^{(2)2}_{1}(t)\frac{e^{mr}\text{Ei}(-3mr)}{4mr}-g^{(2)2}_{1}(t)\frac{e^{-mr}\text{Ei}(-mr)}{4mr},\\
\phi_{1}(t,r)=\frac{36C^{(2)2}_{1}}{m^2r^6}+\frac{3C^{(2)2}_{1}}{r^4}+\frac{3m^2C^{(2)2}_{1}}{2r^2}-4g^{(2)2}_{1}(t)\frac{e^{-2mr}}{m^6r^6}-8g^{(2)2}_{1}(t)\frac{e^{-2mr}}{m^5r^5}-23g^{(2)2}_{1}(t)\frac{e^{-2mr}}{3m^4r^4}\\
\phantom{\phi_{1}(t,r)=}-14g^{(2)2}_{1}(t)\frac{e^{-2mr}}{3m^3r^3}-g^{(2)2}_{1}(t)\frac{e^{-2mr}}{2m^2r^2}+\frac{3C^{(2)2}_{1}m^3e^{mr}\text{Ei}(-mr)}{4r}-\frac{3C^{(2)2}_{1}m^3e^{-mr}\text{Ei}(mr)}{4r}\\
\phantom{\phi_{1}(t,r)=}+3g^{(2)2}_{1}(t)\frac{e^{mr}\text{Ei}(-3mr)}{4mr}-3g^{(2)2}_{1}(t)\frac{e^{-mr}\text{Ei}(-mr)}{4mr},\\
\phi_{2}(t,r)=3g^{(2)2}_{1}(t)\frac{e^{-2mr}}{2m^2r^2}+3g^{(2)2}_{1}(t)\frac{e^{mr}\text{Ei}(-3mr)}{4mr}-3g^{(2)2}_{1}(t)\frac{e^{-mr}\text{Ei}(-mr)}{4mr}.
\end{array}\right.\tag{52}
\end{equation}
\end{widetext}
The exponential integral $\text{Ei}(x)$ is defined as
\begin{displaymath}
\text{Ei}(x)=\int^{x}_{-\infty}\frac{e^{y}}{y}dy,
\end{displaymath}
where when $x>0$, the integral has to be understood in terms of the Cauchy principal value due to the singularity of the integrand at zero. $g^{(4)}_{1}(t)$ is the function of time $t$, and its dimension is length$^{-1}$. By
(\ref{49})--(\ref{52}), we know that there are four contributions to $g^{(4)}_{tt}$, $g^{(4)}_{rr}$, and $R^{(4)}$ in the above Taylor expansion of $f$. The first two ones are the Ricci scalar and its quadratic term, which are in $\phi_{tt}(t,r)$, $\phi_{rr}(t,r)$, and $\phi(t,r)$. The third one is the term $R\mathcal{G}$, which is in $\phi_{tt1}(t,r)$, $\phi_{rr1}(t,r)$, and $\phi_{1}(t,r)$. The fourth one is the cubic term of the Ricci scalar, which is in $\phi_{tt2}(t,r)$, $\phi_{rr2}(t,r)$, and $\phi_{2}(t,r)$.

If we impose an additional
condition of the standard post-Newtonian gauge~\cite{21,31} to
simplify above calculation, we find that the results at $O(2)$
order lose the Yukawa-like characteristic and the oscillting-like
characteristic, i.e., the results at $O(2)$ order reduce down to
the perturbed version of standard Schwarzschild solution at $O(2)$
order. Because $f(R,\mathcal{G})$ gravity and $f(R)$ gravity have
the same spatial behavior at $O(2)$ order, we can find the proof
in Ref.~[21] to support this conclusion. In this reference, the
authors adopted the standard post-Newtonian gauge to calculate the
Newtonian limit and the post-Newtonian limit of $f(R)$ gravity,
and (49) implies that the corresponding result at $O(2)$ order is
the same as the perturbed version of standard Schwarzschild
solution at $O(2)$ order. In order to calculate the result at
$O(4)$ order by using the Yukawa-like solution (38), we do not
impose the standard post-Newtonian gauge, and then we obtain
Eq.~(\ref{48}) and its solution (\ref{51}).

$\alpha_{1}$ and $\alpha_{2}$ are two free parameters in (\ref{51}), and their values decide the behaviors of $g^{(4)}_{tt}$, $g^{(4)}_{rr}$, and $R^{(4)}$. Moreover, by (\ref{25}) and (\ref{49}), we know that $\alpha_{1}$ and $\alpha_{2}$ are linked to the term $R\mathcal{G}$ and the term $R^{3}$ respectively in the above Taylor expansion of $f$.  Because $f_{12}=0$ in (\ref{25}) for $f(R)$ gravity, $\alpha_{1}$ is the specific free parameter for $f(R,\mathcal{G})$ gravity by (\ref{49}). If we set $f_{12}=0$ in (\ref{25}), by the definition of $\alpha_{2}$ in (\ref{49}) and the definition of $\mu$ in Ref~\cite{23}, we obtain $\alpha_{2}=-\frac{1}{3\mu^{4}}$, so $\alpha_{2}$ is the common free parameter for $f(R,\mathcal{G})$ gravity and $f(R)$ gravity.

For a gravitational field generated by a ball-like source described by (\ref{23}), as is mentioned above, we just need to let $C^{(2)}_{1}=r_{g}$ in (\ref{51}) and (\ref{52}), and then we can obtain its corresponding results.

Now we discuss a kind of submodel of $f(R,\mathcal{G})$ gravity whose term $R^2$ disappears in the above Taylor expansion of $f$, namely $f_{11}=0$ in (\ref{25}). For this model, we should let $m\rightarrow\infty$ in (\ref{51}) and (\ref{52}) by (\ref{37}), and then we can obtain its correction to the result at $O(2)$ order, namely
\begin{equation}\label{53}
\left\{\begin{array}{l}
g^{(4)}_{tt}(t,r)=-\alpha_{1}\frac{12C^{(2)2}_{1}}{r^6},\\
g^{(4)}_{rr}(t,r)=-\frac{C^{(2)2}_{1}}{r^2}+\alpha_{1}\frac{72C^{(2)2}_{1}}{r^6},\\
R^{(4)}(t,r)=-\alpha_{1}\frac{1080C^{(2)2}_{1}}{r^8},
\end{array}\right.\tag{53}
\end{equation}
which shows that $g^{(4)}_{tt}$, $g^{(4)}_{rr}$, and $R^{(4)}$ are only related to the term $R\mathcal{G}$ other than the Ricci scalar in the above Taylor expansion of $f$ by (\ref{25}) and (\ref{49}). In addition, we should note that for this model, the result at $O(2)$ order is not (\ref{38}) but is its rudimentary form, namely
\begin{equation}\label{54}
g^{(2)}_{tt}(t,r)=-\frac{C^{(2)}_{1}}{r},\quad g^{(2)}_{rr}(t,r)=-\frac{C^{(2)}_{1}}{r},\quad R^{(2)}(t,r)=0,\tag{54}
\end{equation}
which show that (\ref{38}) has lost its Yukawa-like characteristic and reduces down to
the perturbed version of standard Schwarzschild solution (\ref{40}) at $O(2)$ order for the gravitational field generated by a ball-like source described by (\ref{23}). If the model mentioned above is the submodel of $f(R)$ gravity, by (\ref{49}) and (\ref{53}), let $f_{12}=0$, and then there is
\begin{equation}\label{55}
\left\{\begin{array}{l}
g^{(4)}_{tt}(t,r)=0,\\
g^{(4)}_{rr}(t,r)=-\frac{C^{(2)2}_{1}}{r^2},\\
R^{(4)}(t,r)=0,
\end{array}\right.\tag{55}
\end{equation}
which is the perturbed version of standard Schwarzschild solution (\ref{50}) at $O(4)$ order for the gravitational field generated by a ball-like source described by (\ref{23}). To sum up, we draw a conclusion: If the gravitational field is generated by a ball-like source described by (\ref{23}), for the submodel of $f(R)$ gravity whose term $R^2$ disappears in the Taylor expansion of $f$ around a vanishing value of $R$, its general vacuum solutions up to $O(4)$ order in spherically symmetric background are the same with those in GR. This conclusion and (\ref{53}) show that for $f(R,\mathcal{G})$ gravity and $f(R)$ gravity, term $R^{2}$ in corresponding Lagrangian density plays a key role in the WFSM limit up to $O(4)$ order.
\begin{figure}%[t!]
\centering
\includegraphics[scale=0.67]{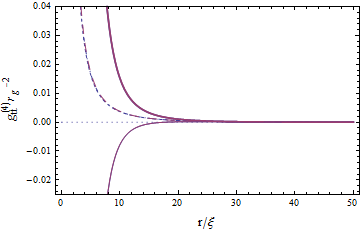}
\caption{\label{fig:9} Plot of the spatial behaviors of $g^{(4)}_{tt}$ for $f(R,\mathcal{G})$ gravity, $f(R)$ gravity, the Starobinsky gravity, and GR with $n$=0.5 and $m$=0.1. The dotted line is the behavior of GR; the dot-dashed curve is the behavior of the Starobinsky gravity; the small dashed curve and the large dashed curve are the behaviors of $f(R)$ gravity with $\alpha_{2} m^4=-0.5,\ 0.5$ respectively; the thick solid curve is the behaviors of $f(R,\mathcal{G})$ gravity with $\alpha_{2} m^4=-0.5,\ 0.5$ and $\alpha_{1} m^4=0.5$; the thin solid curve is the behaviors of $f(R,\mathcal{G})$ gravity with $\alpha_{2} m^4=-0.5,\ 0.5$ and $\alpha_{1} m^4=-0.5$.}
\end{figure}

\begin{figure}%[b!]
\centering
\includegraphics[scale=0.67]{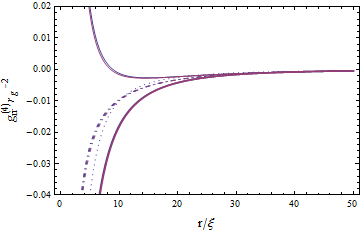}
\caption{\label{fig:10} Plot of the spatial behaviors of $g^{(4)}_{rr}$ for $f(R,\mathcal{G})$ gravity, $f(R)$ gravity, the Starobinsky gravity, and GR with $n$=0.5 and $m$=0.1. The dotted curve is the behavior of GR; the dot-dashed curve is the behavior of the Starobinsky gravity; the small dashed curve and the large dashed curve are the behaviors of $f(R)$ gravity with $\alpha_{2} m^4=-0.5,\ 0.5$ respectively; the thick solid curve is the behaviors of $f(R,\mathcal{G})$ gravity with $\alpha_{2} m^4=-0.5,\ 0.5$ and $\alpha_{1} m^4=0.5$; the thin solid curve is the behaviors of $f(R,\mathcal{G})$ gravity with $\alpha_{2} m^4=-0.5,\ 0.5$ and $\alpha_{1} m^4=-0.5$.}
\end{figure}

\begin{figure}%[t!]
\centering
\includegraphics[scale=0.67]{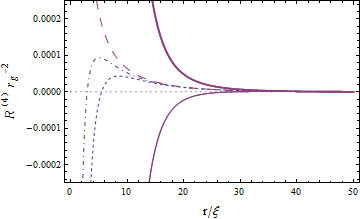}
\caption{\label{fig:11} Plot of the spatial behaviors of $R^{(4)}$ for $f(R,\mathcal{G})$ gravity, $f(R)$ gravity, the Starobinsky gravity, and GR with $n$=0.5 and $m$=0.1. The dotted line is the behavior of GR; the dot-dashed curve is the behavior of the Starobinsky gravity; the small dashed curve and the large dashed curve are the behaviors of $f(R)$ gravity with $\alpha_{2} m^4=-0.5,\ 0.5$ respectively; the thick solid curve is the behaviors of $f(R,\mathcal{G})$ gravity with $\alpha_{2} m^4=-0.5,\ 0.5$ and $\alpha_{1} m^4=0.5$; the thin solid curve is the behaviors of $f(R,\mathcal{G})$ gravity with $\alpha_{2} m^4=-0.5,\ 0.5$ and $\alpha_{1} m^4=-0.5$.}
\end{figure}

Because equations (\ref{48}) are only related to $f_{1}$, $f_{11}$, $f_{12}$, and $f_{111}$, by (\ref{25}), if $f_{12}=0$ and $f_{12}=f_{111}=0$, (\ref{48}) are the differential equations at $O(4)$ order of $f(R)$ gravity and the Starobinsky gravity~\cite{27} respectively. By (\ref{49}), we just let $\alpha_{1}=0$ and $\alpha_{1}=\alpha_{2}=0$ in (\ref{51}), and then we can obtain corrections to (\ref{38}) for these two models respectively, namely
\begin{align*}
\label{56}&\left\{\begin{array}{l}
g^{(4)}_{tt}(t,r)=\phi_{tt}(t,r)+\alpha_{2} m^{4}\phi_{tt\alpha}(t,r),\\
g^{(4)}_{rr}(t,r)=\phi_{rr}(t,r)+\alpha_{2} m^{4}\phi_{rr\alpha}(t,r),\\
R^{(4)}(t,r)=\phi(t,r)+\alpha_{2} m^{4}\phi_{\alpha}(t,r)
\end{array}\right.\tag{56}
\end{align*}
and
\begin{align*}
\label{57}&\left\{\begin{array}{l}
g^{(4)}_{tt}(t,r)=\phi_{tt}(t,r),\\
g^{(4)}_{rr}(t,r)=\phi_{rr}(t,r),\\
R^{(4)}(t,r)=\phi(t,r).
\end{array}\right.\tag{57}
\end{align*}

For the gravitational field generated by a ball-like source described by (\ref{23}), if $g^{(2)}_{1}(t)=-r_{g}m^{2}$ and $g^{(4)}_{1}(t)=nr_{g}^{2}m^{3}$~\cite{23} in (\ref{51}), (\ref{56}), and (\ref{57}), we obtain the static corrections to (\ref{38}) for $f(R,\mathcal{G})$ gravity, $f(R)$ gravity, and the Starobinsky gravity respectively. The free parameter $n$ above is dimensionless, and its value can not be fixed according to the theoretical condition that (\ref{51}), (\ref{56}), and (\ref{57}) should be de Sitter spacetime with a small constant curvature in the asymptotic region and can recover the perturbed version of standard Schwarzschild solution at $O(4)$ order. Moreover, $m$ is also such kind of free parameter, so the values of $m$ and $n$ need experimental evidence to fix. In order to compare the static results in (\ref{51}), (\ref{56}), and (\ref{57}) with the corresponding one in GR, namely (\ref{50}), for $f(R,\mathcal{G})$ gravity, $f(R)$ gravity, and the Starobinsky gravity, we set $n=0.5$ and $m=0.1$. Then how to make sure the values of free parameters $\alpha_{1}$ and $\alpha_{2}$ is a crucial point to obtain the right behavior. By (\ref{37}) and (\ref{49}), there are
\begin{displaymath}
|\alpha_{1} m^{4}|=m^2\left|\frac{f_{12}}{3f_{11}}\right|,\qquad |\alpha_{2} m^{4}|=m^2\left|\frac{f_{111}}{3f_{11}}\right|.
\end{displaymath}
Moreover according to the mathematical interpretation of the WFSM limit, one has
\begin{displaymath}
|f_{111}|<|f_{11}|,\qquad |f_{12}|<|f_{11}|,
\end{displaymath}
so there are
\begin{equation}\label{58}
0<|\alpha_{1} m^{4}|<1,\qquad 0<|\alpha_{2} m^{4}|<1.\tag{58}
\end{equation}
We still need experimental evidence to fix the values of $\alpha_{1}$ and $\alpha_{2}$ further. As an example, we can choose $\alpha_{2} m^4=-0.5,\ 0.5$ and $\alpha_{1} m^4=-0.5,\ 0.5$ in (\ref{51}) by (\ref{58}). Thus,
for these three models above, their spatial behaviors of $g^{(4)}_{tt}$, $g^{(4)}_{rr}$, and $R^{(4)}$ are shown in Figs.~\ref{fig:9}--\ref{fig:11}. From these figures, we know that these behaviors are insensitive to the change of $\alpha_{2} m^4$ which is induced by the cubic term of $R$ in the above Taylor expansion of $f$, so the dot-dashed curve and dashed curves almost coincide in Fig.~\ref{fig:9} and Fig.~\ref{fig:10}. The same reason can be used to explain why there are almost only two solid curves in Figs.~\ref{fig:9}--\ref{fig:11}, and there should be four ones in fact. In addition, these figures show that these behaviors are sensitive to the sign of $\alpha_{1} m^4$ which is induced by the term of $R\mathcal{G}$ in the above Taylor expansion of $f$, so there are two branches for the solid curve.
\section{Conclusions}
In this paper, we develop the WFSM limit of $f(R,\mathcal{G})$ gravity in spherically symmetric background up to $O(4)$ order by generalizing the formalism in Refs.~\cite{21,22}. After having considered the Taylor expansion of a general function $f$ around around vanishing values of $R$ and $\mathcal{G}$, we obtain general vacuum solutions up to $O(4)$ order in a pure perturbative framework. These solutions are time-dependent, and the time-dependent evolution depends on the order of perturbations. Moreover, these solutions depend strictly on the coupling parameters appearing indirectly in the Lagrangian of the theory, namely the partial derivatives of $f$ at $R=0$ and $\mathcal{G}=0$.

A detailed discussion is developed for these solutions in our paper. Compared with the conclusions in Refs.~\cite{21,22}, the solutions at $O(2)$ order show that both $f(R,\mathcal{G})$ gravity and $f(R)$ gravity have the same spatial behaviors, namely the Yukawa-like behavior and the oscillating-like behavior, and the latter is complex valued and is not asymptotic de Sitter spacetime with a small constant curvature in general. Moreover, for the gravitational field generated by a ball-like source, we show these two behaviors related to the $g_{tt}$ components provide two kinds of corrected gravitational potentials about the Newtonian one. Furthermore, for such gravitational field we present its two real valued static behaviors and compared them with the one in GR.

The Newtonian limit of the most general fourth-order theory of gravity, namely $F(X,Y,Z)$ gravity, has been studied in Ref.~\cite{25}, in which the static Yukawa-like behavior has two characteristic lengths for the gravitational field generated by a ball-like source. However for $f(R,\mathcal{G})$ gravity, our solution at $O(2)$ order shows that its static Yukawa-like behavior has only one characteristic length. We show that this is the consequence of the definition of the GB invariant $\mathcal{G}$, so our result about the Yukawa-like behavior is compatible with the corresponding one in $F(X,Y,Z)$ gravity. In addition, for such gravitational field, we indicate that the gravitational potential has the divergency at the position of source, and this is different from the corresponding conclusion of $F(X,Y,Z)$ gravity in Ref.~\cite{25}.

At $O(3)$ order, we show that although the corresponding differential equation is related to time $t$, it could not fix the time-dependent evolution of the two results at $O(2)$ order. We calculate the correction to the Yukawa-like behavior at $O(2)$ order up to $O(4)$ order, and by this correction and previous results, we draw a conclusion: If the gravitational field is generated by a ball-like source, for the submodel of $f(R)$ gravity whose term $R^2$ disappears in the Taylor expansion of $f$ around a vanishing value of $R$, its general vacuum solutions up to $O(4)$ order in spherically symmetric background are the same with those in GR. At last, we present the static corrections of such gravitational field to the Yukawa-like behavior for $f(R,\mathcal{G})$ gravity, $f(R)$ gravity, and the Starobinsky gravity~\cite{27}, and compare them with the one in GR.

Besides, for $f(R,\mathcal{G})$ gravity, such a class of theory have free parameters which should be fixed by experimental evidence. In Ref~\cite{23}, we know that $f(R)$ gravity seems to be a good candidate to explain several data of modern astrophysics and cosmology. Taking into account the results presented here, we know the $f(R,\mathcal{G})$ gravity includes the corresponding results for $f(R)$ gravity in general, together with additional corrected terms determined by some free parameters. Thus, compared with $f(R)$ gravity, $f(R,\mathcal{G})$ gravity might have more freedom to confront with phenomenological data in the future.

\acknowledgments{This work is supported by the National Natural Science Foundation of China (Grants No.~11120101004 and No.~11475006). We also thank Lijing Shao and Xiangdong Zhang for their useful suggestions.}

\end{document}